\documentclass[11pt,twoside,a4paper]{article}
\usepackage[utf8]{inputenc}
\usepackage{latexsym,amsthm}
\usepackage{graphics,physics}
\usepackage{graphicx}
\usepackage{epsfig}
\usepackage{multirow}
\usepackage{amsmath,amssymb,bm}
\usepackage{hhline}
\usepackage{hyperref}
\usepackage{enumitem}
\usepackage{amsfonts}
\usepackage{xcolor}
\usepackage{soul}
\usepackage[font=small,labelfont=bf]{caption}
\usepackage[compat=1.1.0]{tikz-feynman}
\usepackage{subcaption}
\usepackage{hhline}
\usepackage{bm,ulem}
\usepackage{subcaption}
\usepackage{float}
\restylefloat{table}
\usepackage[margin=1in]{geometry}
\usepackage[compat=1.1.0]{tikz-feynman}
\usepackage{braket}
\usepackage{nccmath}
\usepackage{cite}
\usepackage{multirow}
\usepackage{bigcenter}
\usepackage{makecell}
\def\plusheight{-\the\dimexpr\fontdimen22\textfont2\relax}

\title{\bf Asymptotic analysis of  Feynman diagrams and their maximal cuts}

\author{\bf B. Ananthanarayan, \bf Abhijit B. Das, \bf Ratan Sarkar} 

\date{%
    Centre for High Energy Physics,\\ Indian Institute of Science,\\ Bangalore-560012, Karnataka, India }

\begin{document}
\maketitle

\begin{abstract}
The ASPIRE program, which is based on the Landau singularities and the method of Power Geometry to unveil the regions required for the evaluation of a given Feynman diagram asymptotically in a given limit, also allows for the evaluation of scaling coming from the top facets. In this work, we relate the scaling having equal components of the top facets of the Newton polytope to the maximal cut of given Feynman integrals. We have therefore connected two independent approaches to the analysis of Feynman diagrams.

\end{abstract}

\section{Introduction}
The present work is a sequel to Ref.\cite{anant} which presents a novel approach to the Method of Regions\cite{Beneke:1997zp,Smirnov:1999bza, Smirnov:1998vk, Smirnov:2001in, Smirnov:2002pj,Smirnov:1994tg} (MoR) based on the analysis of Landau equations associated with given Feynman diagrams. The algorithm also allows us to compute the scalings of `top facets' which in this work are related in some cases to the maximal cuts of these Feynman diagrams, thereby allowing us to study generalized unitarity in a novel manner to be further explained below.\\
The description of elementary particle physics through perturbative quantum field theory has been very successful. One expresses the field theoretical amplitudes as an expansion in Feynman integrals. The calculation of Feynman integrals with various scales is very difficult. One needs to, very often, calculate higher order loop corrections to these multi-scale Feynman integrals, in order for having better predictions for the field theoretical observables.\\
MoR is one of the useful methods for the evaluation of the multi-scale Feynman integrals. This method uses the hierarchies between various scales of the problem to construct a small expansion parameter, performs Taylor expansion in each of the regions, and evaluates the integral in each of the regions. The final result consists of the sum of the contributions coming from all the regions.\\ 
MoR had been successfully applied in many examples. The foundation and generalization of MoR had been discussed in Ref.\cite{Jantzen:2011nz}. Very recent progress towards the proof of MoR can be found in Ref.\cite{Semenova:2018cwy}, where Lee-Pomeransky representation of Feynman integrals\cite{Lee:2013hzt} had been used to describe MoR. In another recent work\cite{Mishima:2018olh}, MoR had been employed in a systematic way to evaluate two-loop non-planar diagram appearing in the Higgs pair production cross-section at the next-to-leading order.\\
The identification of the regions for a multi-scale multi-loop Feynman integral in a given limit is a non-trivial task. The automatic identification of the regions based on geometrical approach can be found in Ref.\cite{pak_smirnov}. The program had been named as ASY. The potential and Glauber regions were undetected in the first version. This issue had been fixed in Ref.\cite{jantzen_smirnov}. ASY had been implemented inside FIESTA\cite{Smirnov:2015mct} to reveal the regions and numerically evaluate the expansion of the given integral with certain accuracy.\\
The Mathematica program ASPIRE\cite{anant} is based on an alternative formalism, which also unveils the regions associated with a given multi-scale Feynman integral in a given limit. The construction begins with the finding of the sum of Symanzik polynomials of first and second kind. One then finds the Gr\"obner basis of the Landau equations. By mapping the Gr\"obner basis elements to the origin, co-ordinate axes, co-ordinate planes, one obtains a set of linear transformations. All the transformations are applied to the sum of the Symanzik polynomials, which are then analyzed within the framework ``Power Geometry"\cite{bruno:book,bruno_batkhin,bruno2015}. For all of the obtained polynomials, one finds the support of the corresponding polynomials. The convex hull of the support then gives the Newton polytope. One finds the normal vector for the facets of the polytopes based on certain conditions. The set of unique normal vectors gives the desired set of scalings required for the asymptotic expansion of Feynman integrals in given limits. \\
While analyzing a given Feynman diagram within the framework ASPIRE, two types of facets of the Newton polytopes had been considered. For a given sum(polynomial), the bottom facet of the Newton polytope is defined to be the facet for which the points other than the vertices of that facet lie above that facet. On the other hand, top facet is the opposite case. The mathematical definitions of bottom and top facet are given in the appendix~\ref{topBottom} and also we give a detailed description of this discussion for a one loop vertex diagram in section~\ref{sec2}.\\ 
The scalings from the bottom facets with the consideration of small expansion parameter lead to the well known case of ``Regions"\cite{pak_smirnov,jantzen_smirnov,anant}. In this work, we explore the complementary case, i.e., we consider top facet scaling with the freedom of choosing the expansion parameter to be large. The set of Landau equations\cite{Landau:1959fi, Eden:1966dnq} for a given Feynman integral while combined with Bruno's theorem\cite{bruno:book,bruno_batkhin,bruno2015} in Power Geometry implies that the top facet scalings with equal components correspond to the case of maximal cut of the given integral. We explore the correspondence between the parametric integrals constructed based on the scalings having equal components of the top facets and the maximal  cut for given Feynman diagrams.\\
The discontinuity due to the Landau singularities is given in terms of cut Feynman diagrams, by replacing the Feynman propagators by delta functions\cite{Cutkosky:1960sp}. A Feynman diagram is said to be maximally cut when all of its propagators are replaced by Dirac-$\delta$ functions, i.e., all the internal lines are put on-shell.\\ 
The cut Feynman integrals had been studied in a series of works\cite{Abreu:2017ptx, Abreu:2015zaa, Abreu:2017enx, Abreu:2014cla, Harley:2017qut,Frellesvig:2017aai,Abreu:2017mtm,Lee:2012te,Klemm:2019dbm}. These studies show various mathematical structures of the cut Feynman integrals. In Refs.\cite{Abreu:2014cla,Abreu:2017mtm}, some conjectures on the relation of these cut integrals with co-products of multiple polylogarithms in Hopf algebra give an interesting way to compute original Feynman integrals without doing actual integration, but evaluating the comparatively easier cut integral. This method actually relies on the possibility of expressing the original Feynman integral and the cut Feynman integral in terms of multiple polylogarithms. In Refs.\cite{Harley:2017qut,Frellesvig:2017aai}, cut Feynman integrals had been evaluated in a systematic approach using Baikov-Lee representation. Maximal unitarity cut has been connected to the dimensional recurrence relations for multi-component integrals in Ref.\cite{Lee:2012te}. In a recent work\cite{Klemm:2019dbm}, maximal cuts of Feynman diagrams have been analyzed based on multi-dimensional residues in a geometric way.\\ 
In this paper, we use the method of residues\cite{Abreu:2017ptx} to evaluate the cut integrals. The main idea is the equivalence of evaluating the original Feynman integral with cut propagators replaced by Dirac-$\delta$ functions and evaluating the integral of the residue of original Feynman integral at the singularities due to cut propagators. The evaluation of the residues involves deforming the integration contour to include the poles or singularities in Leray's multivariate residue calculus. Right now the method of residues has been worked out only on one loop Feynman integrals. The extension for more than one loop case is a future research work. We use the results directly from literature for the one loop cases that we study and for the two loop case we solve the problem in two parts, i.e., evaluating the results for the one loop case and then applying it to solve the two loop problem by directly using the Dirac-$\delta$ functions inside the integral.
\\ 
Lastly, we derive the cut integrals in Feynman parametric form for the one loop case in order to have a study of the correlation between general cuts and the top integrals with unequal scalings as the loop momentum representation of this kind of top integrals is not expressible in a general form.\\
The organization of this paper is the following:\\
In section~\ref{Form}, we review the basics of Power Geometry and discuss the method to obtain the asymptotic solution of a given finite algebraic sum. For a generic Feynman integral, the Feynman parametric form of the integral in terms of Symanzik polynomials has been discussed in section~\ref{para_int}. In section \ref{sec2}, we present brief description of the Mathematica program ASPIRE. In section\ref{sec3}, we discuss the correspondence of the top facet scalings with equal components to the maximal cut of Feynman diagrams. We show the consideration of the limit of large mass is justified in section \ref{timp}. In section \ref{cormctop}, we derive the one loop generalization of correlation of the maximal cut to the top facet integral with equal components using the large mass expansion limit. In section \ref{topint}, we present the generalized one loop formula for the top facet scalings with equal components. Two one loop diagrams and a two loop non-planar diagram have been analyzed in section \ref{sec4}. We conclude in section \ref{sec5}. We present in appendix \ref{appenA} the description of Mathematica notebooks, the comparison of ASPIRE and ASY for the given examples, and the basic mathematical formulae used in this work. In appendix \ref{cutalpha} we give the Feynman parametric form of the cut integrals in the one loop case.

\section{Formalism}\label{Form}
In this section, we review the frameworks which have been considered during the analysis for obtaining the connection between the scalings with equal components of top facets and the maximal cut Feynman integrals. The framework ASPIRE uses Power Geometry\cite{bruno:book,bruno_batkhin,bruno2015} to find the Regions required for the evaluation of Feynman diagrams by expanding asymptotically in each of the Regions. We start this section with the basic definitions used in Power Geometry and the way to get the asymptotic solutions for a given sum (polynomial), analyzed in the framework Power Geometry. 

\subsection{Power Geometry and the asymptotic solutions for a given sum }
Let us consider a finite sum $g(Q)=\sum g_R Q^R$, where $Q=\left(\alpha_1,\alpha_2,\cdots,\alpha_n\right)$, $R=\left(r_1,r_2,\cdots,r_n\right)\in \mathbb{R}^n$ and $g_R$ are the constant coefficients. By $Q^R$, we mean the terms $\alpha_1^{r_1} \alpha_2^{r_2}\cdots \alpha_n^{r_n}$.\\
Below we give few definitions which are necessary, when one deals with the method of Power Geometry.
\begin{itemize}
\item Support of the sum : \\ The set of vector exponents, $R=\left(r_1,r_2,\cdots,r_n\right)\in \mathbb{R}^n$ is called the support $S(g)$ of the given sum $g(Q)=\sum g_R Q^R$.
\item Newton polytope : \\ The convex hull of the support is called the Newton polytope. It consists of generalized facets $\Gamma_j^d$, where $d$ is the dimension of the facets and the label $j$ stands for the $j$-th facet. For our case, we always consider $d=2$.
\item Truncated sum :\\ Each of the generalized facets $\Gamma_j^d$ corresponds to a sum $\hat{g}_j^d=\sum g_R Q^R$, where $R\in \Gamma_j^d \cap s(g)$. $\hat{g}_j^d$ is called the truncated sum.
\item Normal cone : \\ We consider the dual space, $\mathbb{R}_{*}^{n}$ to the space $\mathbb{R}^n$. We define the scalar product $c_j =  \braket{R,S}$, where $R\in \mathbb{R}^n$ and $S\in \mathbb{R}_{*}^{n} $. The set of all points $S$ for which $c_j$ becomes maximum for all the points $R\in \Gamma_j^d$, is called the normal cone of the generalized facet $\Gamma_j^d$. In our case, as we deal with $d=2$, we consider only the outward normal vector to each of the facets.
\item Cone of the problem : \\ The set of points, $S\in \mathbb{R}_{*}^{n} $ such that the curves of the form of the eq.(\ref{thereom}) that fill the space$(\alpha_1,\alpha_2,\cdots,\alpha_n),$ to be studied is called the cone of the problem.
\end{itemize} 
\indent
{\bf Bruno's Theorem:}\\
If curve
\begin{eqnarray}\label{thereom}
\alpha_1=a_1 x^{s_1}(1+o(1)),\nonumber \\ 
\alpha_2=a_2 x^{s_2}(1+o(1)),\nonumber \\
\vdots \nonumber \\
\alpha_n=a_n x^{s_n}(1+o(1)),\nonumber \\
\end{eqnarray}\\
where $a_i$ and $s_i$ are constants, lie in the set $\mathcal{G}$ as $x \rightarrow \infty$ and the vector $\lbrace s_1, s_2,\cdots, s_n\rbrace \in U^d_j$, then the first approximation $\alpha_1=a_1 x^{s_1}, 
\alpha_2=a_2 x^{s_2},\cdots, \alpha_n=a_n x^{s_n}$ of eq.(\ref{thereom}) satisfies the truncated sum $\hat{g}^d_j=0$.
\\
One wishes to obtain the set $\mathcal{G}=\lbrace Q : g(Q)=0\rbrace$ near singular points $Q=Q_0$, or singular curves $\mathcal{C}$, or singular surfaces $\mathcal{S}$ consisting of the singular points. Below we discuss the steps for obtaining the solution set $\mathit{g}$ for each of the facets of the Newton polytope :
\begin{enumerate}
\item Certain transformations $Q(\alpha_1,\alpha_2,\cdots,\alpha_n)\rightarrow Q'(\alpha_1',\alpha_2',\cdots,\alpha_n')$ need to be performed in order for mapping the singular points, singular curves, and singular surfaces to the origin, co-ordinate axes, and co-ordinate planes respectively.
\item Find $g(Q')$ and the corresponding support $S(g)$.
\item Obtain the Newton polytope for $g(Q')$ and the outward normal vectors $\lbrace s_1,s_2,\cdots,s_n\rbrace$ for each of the facets.
\item The Bruno's theorem\ref{thereom} then gives us the desired solution set $\mathcal{G}$ at the leading order.
\end{enumerate}
\indent
We see that the normal vector for each of the facets of the Newton polytope, resulting from a given sum or polynomial, gives an asymptotic solution according to Bruno's theorem~\ref{thereom}. 
 
\subsection{Parametric representation of Feynman integrals}\label{para_int}
For the sake of completeness, we here briefly discuss parametric representation\cite{Lee:2013hzt,Smirnov:2012gma, Smirnov:2006ry,Weinzierl:2006qs} of a generic Feynman diagram. \\
Consider a Feynman diagram having L loop momenta $(l_1,l_2,...,l_L)$, E external momenta $(p_1,p_2,...,p_E)$ in the generic form,

\begin{equation}\label{eq1}
I(n_1,n_2,..,n_m)=(e^{\epsilon \gamma_E} \mu^{2\epsilon})^L \int \prod_{i=1}^{L}\frac{d^D l_i}{\left(i\pi^{\frac{D}{2}}\right)^{L}}\prod_{\alpha=1}^{m} {D_\alpha}^{-n_\alpha},
\end{equation}
where $D_\alpha= A^{ij}_{\alpha} l_i.l_j+2 B^{ik}_{\alpha} l_i.p_k+C_{\alpha}$ are the given set of propagators. $A, B$ are respectively $L\times L$, $L\times E$ matrices and $C$ are constants. The parameter $\mu$ is arbitrary having mass dimension 1. We put $\mu=1$ throughout our calculations.\\
One can express eq.(\ref{eq1}) in the following form,
\begin{equation}\label{par_int}
\begin{split}
I(n_1,n_2,...,n_m)=(e^{\epsilon \gamma_E} )^L  \frac{\Gamma((n_1+n_2+...+n_m)-\frac{Ld}{2})}{\prod_{\alpha}n_{\alpha}}\int \prod_{\alpha} dz_{\alpha} {z_{\alpha}}^{n_{\alpha}-1}\delta(1-\sum_{\alpha}z_{\alpha}) & \\ \times \frac{{\mathcal{F}}^{(\frac{LD}{2}-(n_1+n_2+...+n_m))}}{{\mathcal{U}}^{(\frac{(L+1)D}{2}-(n_1+n_2+...+n_m))}},
\end{split}
\end{equation}  
where $\mathcal{U}$ and $\mathcal{F}$ are the Symanzik polynomials, of degree $L$ and $L+1$ respectively.\\
In this work, we use the parametric representation for a generic Feynman diagram to construct the integrals based on certain scalings, coming from the top facets (eq.(\ref{top})) of the Newton polytopes. 
\subsection{The Mathematica program - ASPIRE}\label{sec2}
The Mathematica program ``ASPIRE" had been developed to isolate the regions associated with multi-scale, multiloop Feynman diagrams in a given kinematic limit. The formalism of ASPIRE is based on the consideration of singularities of the given Feynman integral and the associated Landau equations and analysis of the sum of the Symanzik polynomials of first and second kind using Power Geometry.\\ 
The program ASPIRE has the following steps:
\begin{enumerate}
\item For a given multi-scale Feynman integral in a given limit, find the Symanzik polynomials $\mathcal{U},\mathcal{F}$.
\item Find the Gr{\"o}bner basis of the Landau equations$\lbrace F, \frac{\partial F}{\partial \alpha_i}\rbrace$, where $\alpha_i $ are the alpha parameters.
\item Map the Gr\"obner basis elements to origin, co-ordinate axes, coordinate planes via linear transformations.
\item Construct $\mathcal{G=U+F}$ polynomials under the consideration of the obtained linear transformations, as mentioned in the previous step.
\item Find the support of each of the $\mathcal{G}$ polynomials.
\item Find the convex hull of the obtained support. Thus one obtains the Newton polytopes.
\item Look for the normal vectors corresponding to each of the facets of the Newton polytopes.
\item The set of the components of the valid normal vectors then gives the set of desired regions.
\end{enumerate}
If for a given sum, one constructs Newton polytope with the vector exponents $\vec{r}$, and $\vec{v}$ is the outward normal vector to the facets of the polytope, then bottom facets of the Newton polytope are those facets which satisfy the following conditions,
\begin{equation}\label{bottom}
\begin{cases}
\vec{r}.\vec{v}= c & \text{for the points on the facets.}\\
\vec{r}.\vec{v}> c & \text{for the points which lie above the facets.}
\end{cases}
\end{equation} \\ 
The top facets of the Newton polytope are defined as,
\begin{equation}\label{top}
\begin{cases}
\vec{r}.\vec{v}= c & \text{for the points on the facets.}\\
\vec{r}.\vec{v}< c & \text{for the points which lie below the facets.}
\end{cases}
\end{equation}
\\
It is important to note that we consider the expansion parameter $x$ to be small $\left( \text{i.e.} \quad x\rightarrow 0\right)$ while we consider the analysis for finding the scalings from the bottom facets of the Newton polytope. In the case of top facets, we choose the expansion parameter to be large $\left( \text{i.e.} \quad x\rightarrow \infty \right)$. \\
It is well known that the scalings coming from bottom facets are the regions which are required for the asymptotic expansion of the Feynman integrals in the given limit.\\
Below we consider a one loop vertex diagram considered in \cite{Beneke:1997zp} as an example to demonstrate the above discussion:

\begin{figure}[h]
\centering
  \includegraphics[width=40mm,scale=0.4]{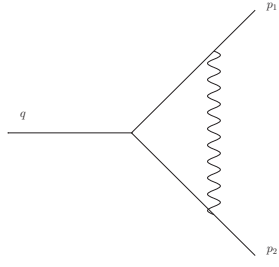}
  \caption{A one loop vertex diagram . The internal solid lines have mass m and the wavy line is massless. }
  \label{toy}
\end{figure}
\indent \\
Fig.\ref{toy} corresponds to the following Feynman integral,
\begin{equation}
I(q^2,m^2)=\int \frac{d^D k}{i \pi^{D/2}} \frac{1}{\left(\left(k+\frac{q}{2}\right)^2-m^2\right)\left(\left(k-\frac{q}{2}\right)^2-m^2\right)\left(k-p\right)^2}
\end{equation}\\
In this case, we have $q=p_1+p_2$, $p=\frac{p_1+p_2}{2}$ and two kinematic invariants $q^2$ and $m^2$. We construct expansion parameter $x=m^2-\frac{q^2}{4}$ to expand the integral in terms, which have certain power in $x$.\\
We find the Symanzik polynomials $\mathcal{U,F}$ using the Mathematica code UF.m\cite{uf} with the following command,
\begin{align}
\text{UF}[ \lbrace k \rbrace,\lbrace -( (k+\frac{q}{2} )^2-m^2 ),-( ( k-\frac{q}{2})^2-m^2 ),-(k-p)^2 \rbrace, \lbrace q^2\rightarrow qq,p q\rightarrow 0, & \nonumber \\ p^2\rightarrow x,m^2\rightarrow x+\frac{qq}{4} \rbrace ],
\end{align}
which gives,
\begin{align}
&\mathcal{U}=\alpha_1+\alpha_2+\alpha_3\\
&\mathcal{F}=\frac{\text{qq} }{4}\alpha_1^2-\frac{1}{2} \text{qq} \alpha_1 \alpha_2+\frac{\text{qq} }{4} \alpha_2^2+x \alpha_1^2+2 x \alpha_2 \alpha_1+x \alpha_2^2,
\end{align} 
where $\alpha_1, \alpha_2$ and $\alpha_3$ are the alpha parameters.\\
We now find the Landau equations, encoding the location of singularities of the integral, 
\begin{align}
&\mathcal{F}=0,\\
&\frac{\partial \mathcal{F}}{\partial \alpha_i}=0, \quad \text{where} \quad i=1,2,3.
\end{align}\\
The Gr\"obner basis of the Landau equations are, \[\left\{\text{qq} x \alpha_2,x \left(\alpha_1+\alpha_2\right),\text{qq} \left(\alpha_1-\alpha_2\right)\right\}\]
We map the Gr\"obner basis elements to the origin, co-ordinate axes with the following transformations,
\begin{eqnarray}
\text{T1}=\left\{\alpha_1\to \alpha_1,\alpha_2\to \alpha_2,\alpha_3\to \alpha_3\right\}\\
\text{T2}=\left\{\alpha_1\to \alpha_1+\frac{\alpha_2}{2},\alpha_2\to \frac{\alpha_2}{2},\alpha_3\to \alpha_3\right\}\\
\text{T3}=\left\{\alpha_1\to \frac{\alpha_1}{2},\alpha_2\to \frac{\alpha_1}{2}+\alpha_2,\alpha_3\to \alpha_3\right\}
\end{eqnarray} 
\indent \\
In this example, we discuss the analysis with the transformation $T1$ only. Analysis with the other two transformations ($T2$ and $T3$) can be found in the ancillary file \texttt{OneloopVertex.nb}. The $\mathcal{G}$ polynomial for the transformation $T1$ is given by, 
\begin{equation}
\mathcal{G}=\alpha_1+ \frac{\text{qq} }{4}\alpha_1^2 + x \alpha_1^2+\alpha_2-\frac{1}{2} \text{qq} \alpha_1  \alpha_2+2 x \alpha_1 \alpha_2+\frac{\text{qq} }{4}\alpha_2^2+x \alpha_2^2+\alpha_3
\end{equation}
We compute the support of $\mathcal{G}$ by extracting the vector exponents of each of the terms,
\begin{align}
S=\left(
\begin{array}{cccc}
 0 & 1 & 0 & 0 \\
 0 & 2 & 0 & 0 \\
 1 & 2 & 0 & 0 \\
 0 & 0 & 1 & 0 \\
 0 & 1 & 1 & 0 \\
 1 & 1 & 1 & 0 \\
 0 & 0 & 2 & 0 \\
 1 & 0 & 2 & 0 \\
 0 & 0 & 0 & 1 \\
\end{array}
\right)
\end{align} 
The co-ordinates of the points are considered in $(x,\alpha_1,\alpha_2,\alpha_3)$-space. We assign label for each of the points of the support $S$ as $\lbrace 1(0,1,0,0),2(0,2,0,0),3(1,2,0,0),\\ 4(0,0,1,0),5(0,1,1,0),6(1,1,1,0),7(0,0,2,0),8(1,0,2,0),9(0,0,0,1)\rbrace$. 
The convex hull of the points of $S$ gives the facets of the Newton polytope,
\begin{align}
NP=\left(
\begin{array}{cccc}
 1 & 2 & 3 & 9 \\
 1 & 2 & 7 & 8 \\
 1 & 2 & 8 & 3 \\
 1 & 2 & 9 & 7 \\
 1 & 3 & 8 & 9 \\
 1 & 4 & 7 & 9 \\
 1 & 4 & 8 & 7 \\
 1 & 4 & 9 & 8 \\
 2 & 3 & 9 & 8 \\
 2 & 7 & 8 & 9 \\
 4 & 7 & 9 & 8 \\
\end{array}
\right)
\end{align}
\\
We now find the normal vector for each of the facets of the Newton polytope with the following considerations:
\begin{enumerate}
\item The component of the normal vector corresponding to the $x$-axis( i.e. zeroth component) should be non-zero.
\item The facets which satisfy eq.(\ref{bottom}) have been labelled as ``$surf\rightarrow -1 $"(i.e. bottom facets). and the facets which satisfy eq.(\ref{top}) have been labelled as ``$surf\rightarrow 1$"(i.e. top facets).
\item One obtains ``Null" when the zeroth component of the normal vector is zero.
\end{enumerate}
\indent
We obtain the following normal vectors corresponding to the facets of $NP$,
\begin{align}
\left(
\begin{array}{ccccccccccc}
 & \text{Null}\\ & \text{Null}\\ & \text{Null}\\ & \{v(1)\to 0,v(2)\to 0,v(3)\to 0,c\to 0,\text{surf}\to -1\}\\ & \{v(1)\to -1,v(2)\to -1,v(3)\to -1,c\to -1,\text{surf}\to 1\}\\ & \{v(1)\to 0,v(2)\to 0,v(3)\to 0,c\to 0,\text{surf}\to -1\} \\ & \text{Null}\\ & \{v(1)\to -1,v(2)\to -1,v(3)\to -1,c\to -1,\text{surf}\to 1\}\\ & \text{Null}\\ & \text{Null}\\ & \text{Null} \\
\end{array}
\right)
\end{align}
We see that with the transformation $T1$, only one region $\lbrace 0,0,0\rbrace$ is isolated. With other two transformations $ T2, T3$, two other regions $\lbrace 1/2,0,0 \rbrace$ and $\lbrace 0,1/2,0 \rbrace$ are recovered. \\ 
There is one more scaling $ \lbrace -1,-1,-1 \rbrace$, which comes from the top facets of the Newton polytope. In this paper, we construct the parametric integral using eq.\ref{par_int} for the top facet scalings having equal components i.e. $\lbrace -1,-1,-1,\cdots,-1 \rbrace$ and find the correspondence of the top facet scalings having equal components to the maximal cut of given Feynman diagrams. 
\subsection{Top facet scaling with equal components and the maximal  cut Feynman diagram}\label{sec3}
Consider a generic Feynman integral,
\begin{align}\label{Feyn}
I(m_i^2,p_i^2)=\int \prod_i \frac{d^Dk_i}{i\pi^{D/2}} \frac{1}{\prod_j (q_j^2-m_j^2)^{n_j}},
\end{align}
where $m_i$ is the mass of i-th internal line, and $p_i$ are the external momenta. The momenta $q_i$ are the linear combination of loop momenta $k_i$ and the external momenta $p_i$.
In Feynman parametric form, eq.~\ref{Feyn} can be written as,
\begin{align}
I(m_i^2,p_i^2)=\int \prod_j d\alpha_j \prod_i \frac{d^Dk_i}{i\pi^{D/2}} \frac{\delta(1-\sum_j \alpha_j)}{ ( \sum_j \alpha_j(q_j^2-m_j^2))^{\sum_j n_j}} 
\end{align}
The Landau singularities are given by,
\begin{equation}
\sum_j \alpha_j (q_j^2-m_j^2)=0
\end{equation}
Each of the facets of the Newton polytope corresponds to an asymptotic solution in the alpha parameter space according to Bruno's theorem \ref{thereom}. We choose the expansion parameter, $x\rightarrow \infty$ for the top facets. This means we are moving far away from the origin.\\
The scalings $\lbrace s_1,s_2,...,s_j\rbrace $ coming from the top facets of the polytopes imply the asymptotic solutions of the form $\lbrace \alpha_1\sim x^{s_1}, \alpha_2\sim x^{s_2},...., \alpha_i\sim x^{s_j}\rbrace$. The scalings can be given a constant shift. If $\vec{S}=\lbrace s_1,s_2,\cdots,s_j \rbrace$ is a scaling coming from one of the facets of the Newton polytope, then $\vec{S'}=\vec{S}+\vec{A}=\lbrace s_1+a,s_2+a,\cdots,s_j+a \rbrace$ corresponds to the same scaling $\vec{S}$.\\
The expansion parameter $x$ being large, the top facet scaling with equal components essentially  gives,\begin{equation}
\alpha_j\neq 0,\quad \text{for all j}
\end{equation}
Thus, for the top facet scaling having equal components, one has
\begin{equation}
q_j^2-m_j^2=0, \quad \text{for all j} 
\end{equation}
This is the on-shell condition for the all the internal lines of the given diagram and hence the case of the maximal cut for the given diagram.\\
This analysis motivates us to express the maximally cut diagram in terms of the integrals constructed from the scaling(with equal components) of the top facet of the Newton polytope.
\subsection{An important remark on the top facet scaling $\lbrace -1,-1,\cdots,-1\rbrace$}\label{timp}
The top scaling $\lbrace -1,-1,\cdots,-1 \rbrace $ corresponds to a set of Symanzik polynomials $\mathcal{U}_{t}$, $\mathcal{F}_{t}$ which can be obtained from the original Symanzik polynomials $\mathcal{U,F}$ by simply putting $q^2\rightarrow 0$ with non-zero $m^2$ in the original $\mathcal{U, F}$. This argument has been checked for all of the examples we have considered, and hence the consideration of this top facet in the large-$m^2$ limit is justified.\\
We demonstrate the above conclusion with an example of a two loop self energy diagram,

\begin{equation}
I(m^2,q^2,d)=\int \frac{d^d k_1 d^d k_2}{(i\pi^{d/2})^2}\frac{1}{(k_1^2-m^2)((q-k_1)^2-m^2)(k_2^2-m^2)((q-k_2)^2-m^2)((k_1-k_2)^2-m^2)}
\end{equation} \\
The Symanzik polynomials for the integral is,
\begin{align}
& \mathcal{U}=\alpha _1 \alpha _3+\alpha _2 \alpha _3+\alpha _5 \alpha _3+\alpha _1 \alpha _4+\alpha _2 \alpha _4+\alpha _1 \alpha _5+\alpha _2 \alpha _5+\alpha _4 \alpha _5 \\ &
\mathcal{F}=(\alpha _3 \alpha _1^2+\alpha _4 \alpha _1^2+\alpha _5 \alpha _1^2+\alpha _3^2 \alpha _1+\alpha _4^2 \alpha _1+\alpha _5^2 \alpha _1+2 \alpha _2 \alpha _3 \alpha _1+2 \alpha _2 \alpha _4 \alpha _1+2 \alpha _3 \alpha _4 \alpha _1+\nonumber \\ &  2 \alpha _2 \alpha _5 \alpha _1+3 \alpha _3 \alpha _5 \alpha _1+3 \alpha _4 \alpha _5 \alpha _1+\alpha _2 \alpha _3^2+\alpha _2 \alpha _4^2+\alpha _2 \alpha _5^2+\alpha _3 \alpha _5^2+\alpha _4 \alpha _5^2+\alpha _2^2 \alpha _3 \nonumber \\ &+\alpha _2^2 \alpha _4+2 \alpha _2 \alpha _3 \alpha _4+\alpha _2^2 \alpha _5+\alpha _3^2 \alpha _5+\alpha _4^2 \alpha _5+3 \alpha _2 \alpha _3 \alpha _5+3 \alpha _2 \alpha _4 \alpha _5+2 \alpha _3 \alpha _4 \alpha _5) m^2+ \nonumber \\ & \left(-\alpha _1 \alpha _2 \alpha _3-\alpha _1 \alpha _4 \alpha _3-\alpha _2 \alpha _4 \alpha _3-\alpha _2 \alpha _5 \alpha _3-\alpha _4 \alpha _5 \alpha _3-\alpha _1 \alpha _2 \alpha _4-\alpha _1 \alpha _2 \alpha _5-\alpha _1 \alpha _4 \alpha _5\right) q^2
\end{align}
While looking for the $\mathcal{U, F}$ for the top facet$\lbrace -1,-1,\cdots,-1\rbrace$ we consider the limit $m^2$ to be large and $q^2\rightarrow 0$. We have implemented this consideration in the function \texttt{getLOUF} in ASPIRE program. Using this function, we obtain the following Symanzik polynomials for the top facet$\lbrace -1,-1,\cdots,-1\rbrace$,
\begin{align}
& \mathcal{U}_t=\alpha _1 \alpha _3+\alpha _2 \alpha _3+\alpha _5 \alpha _3+\alpha _1 \alpha _4+\alpha _2 \alpha _4+\alpha _1 \alpha _5+\alpha _2 \alpha _5+\alpha _4 \alpha _5 \\ &
\mathcal{F}_t=(\alpha _3 \alpha _1^2+\alpha _4 \alpha _1^2+\alpha _5 \alpha _1^2+\alpha _3^2 \alpha _1+\alpha _4^2 \alpha _1+\alpha _5^2 \alpha _1+2 \alpha _2 \alpha _3 \alpha _1+2 \alpha _2 \alpha _4 \alpha _1+2 \alpha _3 \alpha _4 \alpha _1+\nonumber \\ &  2 \alpha _2 \alpha _5 \alpha _1+3 \alpha _3 \alpha _5 \alpha _1+3 \alpha _4 \alpha _5 \alpha _1+\alpha _2 \alpha _3^2+\alpha _2 \alpha _4^2+\alpha _2 \alpha _5^2+\alpha _3 \alpha _5^2+\alpha _4 \alpha _5^2+\alpha _2^2 \alpha _3 \nonumber \\ &+\alpha _2^2 \alpha _4+2 \alpha _2 \alpha _3 \alpha _4+\alpha _2^2 \alpha _5+\alpha _3^2 \alpha _5+\alpha _4^2 \alpha _5+3 \alpha _2 \alpha _3 \alpha _5+3 \alpha _2 \alpha _4 \alpha _5+2 \alpha _3 \alpha _4 \alpha _5) m^2,
\end{align}
which exactly match with the $\mathcal{U,F}$ in the limit $q^2\rightarrow 0$. We also identify the location of the points on the Newton polytope of $\mathcal{G=U+F}$ which give rise to $\mathcal{U}_{t}$, $\mathcal{F}_{t}$. Thus we confirm that the top facet$\lbrace -1,-1,\cdots,-1 \rbrace$ corresponds to the limit where $q^2$ can be neglected with respect to $m^2$,
\begin{align}
& \mathcal{U}_t=\mathcal{U} \\ &
\mathcal{F}_t=\mathcal{F}|_{q^2\rightarrow 0, m^2\neq 0}
\end{align}
In the following section, we describe the method of evaluating the cut integrals for Feynman diagrams.
\subsection{Brief description of the method of evaluation of cut Feynman diagrams}
In this section, we give a brief review of the recent work\cite{Abreu:2017ptx} for the evaluation of cut Feynman diagrams. We use this method for all of our calculations regarding the evaluation of cuts of Feynman diagrams.\\ 
We start with Leray's Multivariate Residues which states that an integrand (differential form of weight $n$) which is of the form given by, 	
\begin{equation}\label{eq:to_residue}
\omega = \frac{ds}{s^n}\wedge \psi + \theta\,,
\end{equation}
	has residues defined by 
	\begin{equation}\label{defres}
	\textrm{Res}_S[\omega] = \psi_{|S}\,.
	\end{equation}
and the following equation holds
\begin{equation}\label{eq:residue_thm}
	\int_{\delta\sigma}\omega = 2\pi i\,\int_{\sigma}\textrm{Res}_S[\omega]\,.
\end{equation}
where $\wedge$ is the generalization of cross product in higher dimensions, $\psi$ is a differential form of weight $n-1$, $s$ is equivalent to the propagator,  S is the singularity zone, $\sigma \subset S$ and $\delta\sigma$ is the set of points which form a small circle around every point on $\sigma$ but not belonging to S called as the ``Tubular neighborhood" or ``Leray coboundary" which ``wraps around" $\sigma$. \\
One loop Feynman integrals can be written as 
\begin{equation}\label{one-loop}
    I_n^D = \int\omega_n^D\,,
    \end{equation}
    where the integrand is of the form
    \begin{equation}\label{one-loop omega}
    \omega_n^D = \frac{e^{\gamma_{E}\epsilon}}{i\pi^{{D}/{2}}} \frac{d^Dk}{D_1\ldots D_{n}}\,,
\end{equation}
with $D_j = \left(k-q_{j}\right)^2-m_{j}^2+i0$.\\
In order to get the residue of the Feynman integral we have to write the integrand eq.(\ref{one-loop omega}) in the form of eq.(\ref{eq:to_residue}). This can be acheived by a Jacobian transformation from $k$ to $D_j$ after which the integrand can be written in a form
\begin{equation}\label{eq:Baikov}
		\omega_n^D= \frac{2^{-c}\,e^{\gamma_E\epsilon}}{\sqrt{\mu'^c\,H_C}}\,\left(\mu'\,\frac{H_C}{Gram_C}\right)^{(D-c)/2}\,\frac{d\Omega_{D-c}}{i\pi^{D/2}}\,
		\left(\prod_{j \notin C}\frac{1}{D_j}\right)\,\left(\prod_{j\in C}\frac{dD_j}{D_j}\right)\,,
\end{equation}
where the factor $\mu'= (+1)/(-1)$ accounts for the Euclidean/Minkowski space respectively, $C$ is the set of cut propagators and $c$ is the total number of cut propagators, $d\Omega$ is the angular part of the differential $d^Dk$ in the remaining $D-c$ dimensions, $Gram_C$ and $H_C$ are given by
\begin{equation}\label{defgram}
    Gram_C= det((q_i-q_*)\cdot(q_j-q_*))_{i,j\in C\setminus \{*\}},
    H_C= det((q_i-k)\cdot(q_j-k))_{i,j\in C\setminus \{*\}}
\end{equation}
with $\{*\} \in C$.
Thus using eq.(\ref{defres}) gives
\begin{equation}
		\textrm{Res}_C[\omega_n^D] = 2^{-c}\,e^{\gamma_E\epsilon}\,\frac{d\Omega_{D-c}}{i\pi^{D/2}}\,\left[\frac{1}{\sqrt{\mu'^c\,H_C}}\,\left(\mu'\,\frac{H_C}{Gram_C}\right)^{(D-c)/2}\,\left(\prod_{j\notin C}\frac{1}{D_j}\right)\right]_C\,,
		\end{equation}
where the notation $[.]_C$ indicates that the expression inside square brackets should be evaluated on the locus where the cut propagators vanish. \\
	 As discussed earlier, the integral of the residue is actually equivalent to the cut integral and hence we can write the cut integral corresponding to eq.(\ref{one-loop}) as
\begin{align}\label{cut_master}
\mathcal{C}_c I_n= 2^{-c} \frac{(2\pi i)^{[c/2]}e^{\gamma_E \epsilon} }{\sqrt{\mu'^c Y_c}}\left(\mu' \frac{Y_C}{Gram_C}\right)^{(D-c)/2}\int_{S_\perp} \frac{d\Omega_{D-c}}{i \pi^{D/2}}\left[\prod_{j\notin C} \frac{1}{(k-q_j)^2-m_j^2}\right]_C \text{mod } i\pi,
\end{align} 
\begin{equation}\label{defYC}
    \text{where} \quad Y_C=det\left(\frac{1}{2}(-(q_i-q_j)^2+m_i^2+m_j^2)\right)_{i,j\in C}
\end{equation} 
In the following section, we discuss the generalization at one loop for the maximal cut integrals and the correlation with the top facet$\lbrace -1,-1,\cdots,-1 \rbrace$.

\section{A formula of correlation between maximal  cuts and top integral  for the one loop case} \label{cormctop}
\subsection{Unequal masses}
The top integral with scaling $\lbrace -1,-1,\cdots,-1 \rbrace$ simplifies to all $q_i's\to0$ in the original loop momentum representation of top facet integrals. This is because all the $\alpha_i's$ are of equal scaling and the Symanzik polynomials are homogenous in the variables $\alpha_i's$ and hence we just neglect the terms with prefactor $q_i^2$ compared to $m_i^2$.\\
Using eq.(4.10) of \cite{Abreu:2017ptx} we have the one loop Feynman integral given by
\begin{align}\label{eq:uncut_integral}
I_n=&(-1)^n\frac{2^{\sum_{j=0}^{n-2}(D-2-j)}e^{\gamma_{E}\epsilon}}{\pi^{\frac{n-1}{2}}\Gamma\left(\frac{D-n+1}{2}\right)}\int_0^\infty dr^2\frac{\left(r^2\right)^{\frac{D-2}{2}}}{r^2+m_{n-1}^2}
\prod_{j=0}^{n-2}\int_0^1 dt_j \frac{\left[t_j(1-t_j)\right]^{\frac{D-3-j}{2}}}{B_j\left( t_j-T_{j}\right)}\, ,
\end{align}
Since we have all the external momenta equal to zero using eq.(4.8) of \cite{Abreu:2017ptx} we have the top integral
\begin{align}\label{eq:top_integral}
I^{\lbrace -1,-1,\cdots,-1\rbrace}=&(-1)^n\frac{2^{\sum_{j=0}^{n-2}(D-2-j)}e^{\gamma_{E}\epsilon}}{\pi^{\frac{n-1}{2}}\Gamma\left(\frac{D-n+1}{2}\right)}\int_0^\infty dr^2\frac{\left(r^2\right)^{\frac{D-2}{2}}}{r^2+m_{n-1}^2}
\prod_{j=0}^{n-2}\int_0^1 dt_j \frac{\left[t_j(1-t_j)\right]^{\frac{D-3-j}{2}}}{r^2+m_{j}^2}\, ,
\end{align}
The $t_j$ integration is trivial using Beta functions. For the $r^2$ integral first we can do the partial fraction expansion for the denominator and then integrate. Doing this we get
\begin{align}
I^{\lbrace -1,-1,\cdots,-1\rbrace}=&(-1)^n\frac{2^{\sum_{j=0}^{n-2}(D-2-j)}e^{\gamma_{E}\epsilon}}{\pi^{\frac{n-1}{2}}\Gamma\left(\frac{D-n+1}{2}\right)}\sum_{i=0}^{n-1}\frac{\pi\hspace{.1cm}\text{cosec}\left(\frac{D \pi}{2}\right)\left(m_i^2\right)^{\frac{D-2}{2}}}{\left(\prod_{j=0,j\neq i}^{n-1}(-m_i^2+m_{j}^2)\right)}
\prod_{j=0}^{n-2} \frac{\Gamma^2\left(\frac{D-1-j}{2}\right)}{\Gamma\left(D-1-j\right)}\, ,
\end{align}
After using the Legendre duplication formula and some simplification we get,
 \begin{align}\label{actualtop}
I^{\lbrace -1,-1,\cdots,-1\rbrace}=&(-1)^n\pi\hspace{.1cm}\text{cosec}\left(\frac{D \pi}{2}\right)\frac{e^{\gamma_{E}\epsilon}}{\Gamma\left(\frac{D}{2}\right)}\sum_{i=0}^{n-1}\frac{\left(m_i^2\right)^{\frac{D-2}{2}}}{\left(\prod_{j=0,j\neq i}^{n-1}(-m_i^2+m_{j}^2)\right)}
\, ,
\end{align}
This is the general formula for one loop top integrals with unequal masses and scaling having equal components. Using eq.(3.31) of \cite{Abreu:2017ptx} we have the maximal  cut for the one loop integral given by
\begin{align}\label{eq:cutdef-det}
I^{MC}  = 2^{1-n}\frac{\pi^{(D-n+1)/2}(2\pi i)^{\lfloor {n}/{2}\rfloor}e^{\gamma_E\epsilon}}{\Gamma(\frac{D-n+1}{2})\sqrt{\mu'^n Y_n}}\left(\mu' \frac{Y_n}{Gram_n}\right)^{(D-n)/2}
\end{align}
where we have used $c=n$ i.e. the number of cut propagators are equal to the total number of propagators. Here $Gram_n$ and $Y_n$ are the Gram and modified Cayley determinants respectively which are complicated functions of the external propagators and internal masses. So the correlation here is not so obvious as there is no visible proportionality.
\begin{align}\label{eq:topcut}
&I^{MC}  = \frac{\pi^{(D-n+3)/2}(2\pi i)^{\lfloor {n}/{2}\rfloor}\Gamma(D/2)\text{cosec}\left(\frac{D \pi}{2}\right)}{(-2)^{n-1}\Gamma(\frac{D-n+1}{2})\sqrt{\mu'^n Y_n}}\left(\mu' \frac{Y_n}{Gram_n}\right)^{(D-n)/2} \nonumber \\
&  \hspace{4cm}   \times\left[\sum_{i=0}^{n-1}\frac{\left(m_i^2\right)^{\frac{D-2}{2}}}{\left(\prod_{j=0,j\neq i}^{n-1}(-m_i^2+m_{j}^2)\right)}\right]^{-1} I^{\lbrace -1,-1,\cdots,-1\rbrace}
\end{align}
But if we explicitly put the top condition i.e. all $q_i's=0$ in the maximal cut then we are able to get the correlation in form of proportionality. For eq.(\ref{eq:top_integral}) if we evaluate the maximal cut using Cauchy's theorem of sum of residues we get
\begin{align}\label{eq:top1integral}
I^{MC}=&(-1)^n\frac{2^{\sum_{j=0}^{n-2}(D-2-j)}e^{\gamma_{E}\epsilon}}{\pi^{\frac{n-1}{2}}\Gamma\left(\frac{D-n+1}{2}\right)}\sum_{i=0}^{n-1}\frac{\left(-m_i^2\right)^{\frac{D-2}{2}}}{\left(\prod_{j=0,j\neq i}^{n-1}(-m_i^2+m_{j}^2)\right)}
\prod_{j=0}^{n-2} \frac{\Gamma^2\left(\frac{D-1-j}{2}\right)}{\Gamma\left(D-1-j\right)}\, ,
\end{align}
which simplifies to
\begin{align}\label{eq:top2_integral}
I^{MC}=(-1)^n\frac{e^{\gamma_{E}\epsilon}}{\Gamma\left(\frac{D}{2}\right)}\sum_{i=0}^{n-1}\frac{\left(-m_i^2\right)^{\frac{D-2}{2}}}{\left(\prod_{j=0,j\neq i}^{n-1}(-m_i^2+m_{j}^2)\right)}
\, ,
\end{align}
This equation cannot be obtained simply by substituting all $q_i's=0$ in eq.(\ref{eq:cutdef-det}) as it gives a zero Gram determinant in the denominator. So comparing eq.(\ref{actualtop})and eq.(\ref{eq:top2_integral}) we can see a direct correlation in the form of proportionality and hence the correlation equation becomes
 \begin{align}\label{coruneqmass}
     I^{\lbrace -1,-1,\cdots,-1\rbrace}=(-1)^{\frac{D-2}{2}}\pi\hspace{.1cm}\text{cosec}\left(\frac{D \pi}{2}\right)I^{MC}
 \end{align}

\subsection{Equal masses}
Now for the case in which all the masses are equal eq.(\ref{eq:top_integral}) gets modified to
\begin{align}\label{eq:top3_integral}
I^{\lbrace -1,-1,\cdots,-1\rbrace}=&(-1)^n\frac{2^{\sum_{j=0}^{n-2}(D-2-j)}e^{\gamma_{E}\epsilon}}{\pi^{\frac{n-1}{2}}\Gamma\left(\frac{D-n+1}{2}\right)}\int_0^\infty dr^2\frac{\left(r^2\right)^{\frac{D-2}{2}}}{(r^2+m^2)^n}
\prod_{j=0}^{n-2}\int_0^1 dt_j \left[t_j(1-t_j)\right]^{\frac{D-3-j}{2}}\, ,
\end{align}
Integrating this gives
\begin{align}\label{actualtopeqmass}
I^{\lbrace -1,-1,\cdots,-1\rbrace}=(-1)^ne^{\gamma_{E}\epsilon}\left(m^2\right)^{\frac{D-2n}{2}}\frac{ \Gamma \left(n-\frac{D}{2}\right)}{\Gamma (n)}
\end{align}
Now from eq.(\ref{eq:top3_integral}) we will evaluate the maximal cut with top conditions imposed again using the method of residues. 
This time we have a pole of order n, so using Cauchy's theorem for higher order residues, we get,
\begin{align}\label{eq:top4_integral}
I^{MC}=&(-1)^n\frac{2^{\sum_{j=0}^{n-2}(D-2-j)}e^{\gamma_{E}\epsilon}}{\pi^{\frac{n-1}{2}}\Gamma\left(\frac{D-n+1}{2}\right)}\frac{\Gamma(D/2)(-m^2)^{\frac{D-2n}{2}}}{\Gamma(n)\Gamma(\frac{D+2-2n}{2})}
\prod_{j=0}^{n-2}\frac{\Gamma^2\left(\frac{D-1-j}{2}\right)}{\Gamma\left(D-1-j\right)}\, ,
\end{align}
and after simplification we get,
\begin{align}\label{eq:top5_integral}
I^{MC}=&(-1)^n\frac{e^{\gamma_{E}\epsilon}(-m^2)^{\frac{D-2n}{2}}}{\Gamma(n)\Gamma(\frac{D+2-2n}{2})}\, ,
\end{align}
So the correlation equation becomes

\begin{align}\label{coreqmass}
    I^{\lbrace -1,-1,\cdots,-1\rbrace}=(-1)^{\frac{D-2n+2}{2}}\pi \hspace{.1cm} \text{cosec} \left(\frac{\pi  (D-2 n)}{2} \right) I^{MC}
\end{align}

\section{A general formula for the top integrals in the one loop case}\label{topint}
In this section we derive a general formula of top integrals with scaling $\lbrace -1,-1,\cdots,-1 \rbrace$ without using parameterization. As discussed earlier these kind of the top integrals have effectively all external momenta $q_i=0$, we have the following loop momentum representation for this class of top integrals in the one loop case.
\begin{align}\label{onelooptopgen}
    \int\frac{d^Dk}{i\pi^{D/2}}\prod_i\frac{1}{(k^2-m_i^2)^{a_i}}
\end{align}
In order to find a result for this we use the method in Ref.\cite{Suzuki:2002js}. Now consider an Gaussian integral of the form 
\begin{align}\label{exptop}
   I=\int\frac{d^Dk}{i\pi^{D/2}}\text{exp}\left[-\sum_{i=0}^{n-1}\alpha_i(k^2-m_i^2)\right]
\end{align}
Here $\alpha_i's$ are positive parameters. Expanding the exponential function we will get 
\begin{align}\label{expexpand}
  I=\sum_{a_0,a_1,...,a_{n-1}=0}\frac{(-1)^{\sum_{i=0}^{n-1}a_i}}{i\pi^{D/2}}\prod_{i=0}^{n-1}\frac{\alpha_i^{a_i}}{a_i!}\int d^Dk\prod_{j=0}^{n-1}(k^2-m_j^2)^{a_j}
\end{align}
Now using the definition 
\begin{align}\label{defalpha}
    \alpha=\sum_{i=0}^{n-1}\alpha_i
\end{align}
we get eq.(\ref{exptop}) rewritten as 
\begin{align}
    I=\int\frac{d^Dk}{i\pi^{D/2}}\text{exp}\left[-\alpha k^2+\sum_{i=0}^{n-1}\alpha_i(m_i^2)\right]
\end{align}
We can evaluate this Gaussian integral in $D$ dimensions and the result is 
\begin{align}
   I=\frac{1}{i\alpha^{D/2}}\text{exp}\left[\sum_{i=0}^{n-1}\alpha_i(m_i^2)\right]
\end{align}
Now we can expand the exponential to get
\begin{align}\label{expexpand1}
   I=\frac{1}{i\alpha^{D/2}}\sum_{j_0,j_1,...,j_{n-1}=0}\prod_{i=0}^{n-1}\frac{\alpha_i^{j_i}(m_i^2)^{j_i}}{j_i!}
\end{align}
Now if we take the multinomial expansion in $\alpha$ using eq.(\ref{defalpha}) we get
\begin{align}
    \frac{1}{\alpha^{D/2}} =\frac{1}{(\sum_{i=0}^{n-1}\alpha_i)^{D/2}} 
    = \sum_{j_n,j_{n+1},...,j_{2n-1}=0}\Gamma(1-D/2)\prod_{i=0}^{n-1}\frac{\alpha_i^{j_{n+i}}}{j_{n+i}!}
\end{align}
Substituting this in eq.(\ref{expexpand1}) we get
\begin{align}
    I
    = \frac{1}{i}\sum_{j_0,j_1,...,j_{2n-1}=0}\Gamma(1-D/2)\prod_{i=0}^{n-1}\frac{\alpha_i^{j_i+j_{n+i}}(m_i^2)^{j_i}}{j_i!j_{n+i}!}
\end{align}
Now if we put the constraint that $j_i+j_{n+i}=a_i$ then we get 
\begin{align}\label{eqnconstraint}
    I
    =\frac{1}{i} \sum_{a_1,a_2,...,a_{n-1}=0}\prod_{k=0}^{n-1}\alpha_k^{a_k}\sum_{j_0,j_1,...,j_{2n-1}=0}\Gamma(1-D/2)\prod_{i=0}^{n-1}\frac{(m_i^2)^{j_i}}{j_i!j_{n+i}!}
\end{align}
Comparing eq.(\ref{expexpand}) and eq.(\ref{eqnconstraint}), we get
\begin{align}
    \int \frac{d^Dk}{\pi^{D/2}}\prod_{j=0}^{n-1}(k^2-m_j^2)^{a_j}
    = (-1)^{\sum_{i=0}^{n-1}a_i}\prod_{k=0}^{n-1}\Gamma(1+a_k)\sum_{j_0,j_1,...,j_{2n-1}=0}\Gamma(1-D/2)\prod_{i=0}^{n-1}\frac{(m_i^2)^{j_i}}{j_i!j_{n+i}!}
\end{align}
So this is the general result for the top integrals with scaling $\lbrace -1,-1,\cdots,-1 \rbrace $ irrespective of equal or unequal masses. The required results for each case can be derived by analytic continuation of $a_i's$ to their negative values well described in Ref.\cite{Suzuki:2002js}. 

\section{Examples}\label{sec4}
In this section, we evaluate the parametric integral for the top facet scaling with equal components and find their correspondence to the maximal cut for a two point one loop diagram, a three point one loop diagram, and a non-planar two loop diagram.  
\subsection{Two point one loop diagram}
We consider the following integral in dimension $D=4-2\epsilon$,
\begin{equation}
I(q^2,m^2)=e^{\gamma_E \epsilon}\int \frac{d^Dk}{i\pi^{D/2}}\frac{1}{(k^2-m^2)((k-q)^2-m^2)},
\end{equation} 
where $q$ is the external momentum and $m$ is the mass of both internal lines. The expansion parameter is $x=m^2-\frac{q^2}{4}$.

\begin{figure}[h]
\centering
  \includegraphics[width=50mm,scale=0.7]{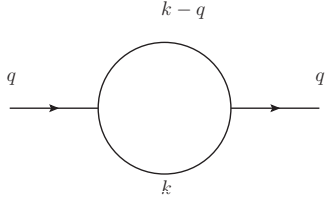}
  \caption{A two point one loop diagram.}
  \label{2.1.dia}
\end{figure}
\noindent
The symanzik polynomials are
\begin{equation}
\mathcal{U}=\alpha_1+\alpha_2
\end{equation}
\begin{equation}
\mathcal{F}=\frac{1}{4} q^2 \alpha_1^2+\frac{1}{4} q^2 \alpha_2^2-\frac{1}{2} q^2 \alpha_1 \alpha_2+x \alpha_1^2+x \alpha_2^2+2 x \alpha_1 \alpha_2,
\end{equation}
where $\alpha_1$ and $\alpha_2$ are the alpha parameters. \\
Using ASPIRE, we find that the above diagram has only one top facet scaling $\lbrace -1,-1 \rbrace$.

\subsubsection{Parametric integral from the top facet scaling $\lbrace -1,-1 \rbrace $ }
We compute the Symanzik polynomials using the top facet scaling $\lbrace -1, -1 \rbrace$,
\begin{eqnarray}
\mathcal{U}=\alpha_1+\alpha_2, \label{Ut1}\\
\mathcal{F}=x(\alpha_1+\alpha_2)^2 \label{Ft1}
\end{eqnarray}
The integral for the scaling $\lbrace
-1,-1\rbrace$ is obtained by substituting the expressions of $\mathcal{U,F}$(eq.(\ref{Ut1}) and eq.(\ref{Ft1})) in eq.(\ref{par_int}),
\begin{align}\label{2.1para}
I^{\lbrace
-1,-1\rbrace}= & e^{\gamma_E \epsilon}\Gamma(2-\frac{D}{2})\int_{0}^{1} d\alpha_1 \int_{0}^{1}d\alpha_2 \frac{\delta(1-\alpha_1-\alpha_2) (\alpha_1+\alpha_2)^{2-D}}{(x(\alpha_1+\alpha_2)^2)^{2-\frac{D}{2}}} \nonumber \\ &
 =e^{\gamma_E \epsilon}\Gamma(\epsilon)\left(m^2-\frac{q^2}{4}\right)^{-\epsilon}
\end{align}
In the limit $m^2>>q^2$, we find
\begin{align}
I^{\lbrace
-1,-1\rbrace}=e^{\gamma_E \epsilon}\Gamma(\epsilon)\left(m^2\right)^{-\epsilon}
\end{align}
This is exactly equal to eq.(\ref{actualtopeqmass}) for n=2.
\subsubsection{The maximal cut integral}
\begin{figure}[h]
\centering
  \includegraphics[width=50mm,scale=0.7]{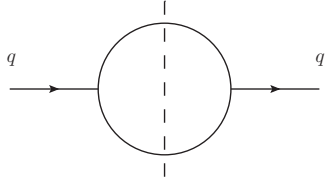}
  \caption{The maximal cut of two point one loop diagram.}
  \label{2.1.cut}
\end{figure}
\noindent The maximal cut for this diagram is obtained by putting both of the two internal lines simultaneously to be on-shell, i.e. we substitute a delta function for both of the propagators. Thus,
\begin{equation}
I^{MC}=e^{\gamma_E \epsilon}\int 
\frac{d^Dk}{i\pi^{D/2}} \delta(k^2-m^2) \delta((k-q)^2-m^2)
\end{equation}
In eq.(\ref{cut_master}) the quantity inside the square bracket is unity because there are no propagators which are not cut for this case and we have $c=n=2$ with $D=4-2\epsilon$ as usual, thus we obtain the maximal cut for the figure.(\ref{2.1.dia}),
\begin{align}\label{cut_2.1}
I^{MC}= i\pi e^{\gamma_E\epsilon}\frac{1}{\sqrt{\mu'^2 Y_C}} \left(\mu' \frac{Y_C}{Gram_C}\right)^{1-\epsilon}\int \frac{d\Omega_{2-2\epsilon}}{i \pi^{2-\epsilon}}
\end{align}
Using eq.(\ref{defgram}) and eq.(\ref{defYC}) for this case we have, \[Gram_C=\begin{vmatrix} q^2 \end{vmatrix}= q^2\] and \[Y_C=\begin{vmatrix} m^2 & -\frac{q^2}{2}+m^2\\-\frac{q^2}{2}+m^2 &  m^2 \end{vmatrix}= \frac{q^2(4m^2-q^2)}{4}\]
Also the angular part of the integration is given by\footnote{This formula is according to the convention followed in \cite{Abreu:2017ptx} which is stated in eq.(\ref{angularint}). }$$ \int d\Omega_{2-2\epsilon} = \frac{2 \pi^{(3-2\epsilon)/2}}{\Gamma((3-2\epsilon)/2)}$$
Thus from eq.(\ref{cut_2.1}), after using the duplication formula of gamma function\footnote{Gamma function duplication formula:\begin{equation*}\label{dup}
    \Gamma(2n)=\frac{1}{\sqrt{\pi}}2^{2n-1}\Gamma(n)\Gamma(n+\frac{1}{2})
\end{equation*}} we obtain the final result for the maximal cut,
\begin{align}\label{2.1MC}
I^{MC}= 2e^{\gamma_E \epsilon} \frac{(4m^2-q^2)^{1-\epsilon}}{\sqrt{q^2(4m^2-q^2)}}\frac{\Gamma(2-\epsilon)}{\Gamma(3-2\epsilon)}
\end{align}

\subsubsection{Correlation between $I^{\lbrace -1,-1\rbrace}$ and $I^{MC}$}
We obtain the following relation: %for $I^{\lbrace -1,-1\rbrace}$ in terms of $I^{MC}$,
%\begin{equation}
%I^{\lbrace -1, -1\rbrace}=f\left(m^2,q^2,\epsilon \right) I^{MC},
%\end{equation}
%where $f\left(m^2,q^2,\epsilon \right)= \frac{4^{\epsilon}}{\pi^{1-\epsilon}} \frac{\Gamma(\epsilon)\Gamma(3-2\epsilon)}{\Gamma(2-\epsilon)}\sqrt{\frac{q^2}{4m^2-q^2}}$.

\begin{equation}
I^{MC}= \frac{2^{1-2\epsilon}\Gamma(2-\epsilon)}{\Gamma(\epsilon)\Gamma(3-2\epsilon)}\left(\frac{4m^2-q^2}{q^2}\right)^{\frac{1}{2}}\quad I^{\lbrace -1,-1 \rbrace}
\end{equation}
Using eq.(\ref{coreqmass}) the correlation for this case is
\begin{align}
    I^{\{-1,-1\}}=(-1)^{1-\epsilon}\pi \hspace{.1cm} \text{cosec} \left(-\pi \epsilon \right) I^{MC}
\end{align}
%\subsubsection{The next-to-maximal -cut integral}
%The next-to-maximal -cut(NMC)  is obtained by cutting one of the two propagators. 
%\begin{equation}
%I^{NMC}=\int d^dk \frac{\delta (k^2-m^2)}{((k-q)^2-m^2)}
%\end{equation} 

\subsection{A one loop scalar triangular diagram}
\begin{figure}[h]
\centering
  \includegraphics[width=50mm,scale=0.4]{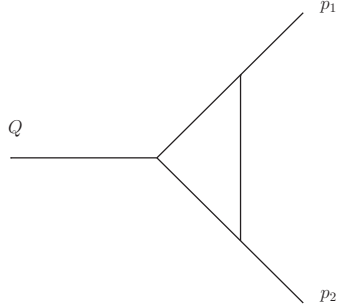}
  \caption{A one loop triangular diagram.}
  \label{triangle}
\end{figure}

We consider the triangular diagram~(\ref{triangle}) in the limit $p_1^2=0$, $p_2^2=0$ and $2p_1.p_2=Q^2$. The integral in this limit is given by,
\begin{equation}
I(Q^2,m^2,D)=\frac{e^{\gamma_E \epsilon}}{i\pi^{D/2}}\int d^Dk \frac{1}{(k^2-2p_1.k)(k^2-2p_2.k)(k^2-m^2)}
\end{equation}\\
The expansion parameter is $\frac{m^2}{Q^2}$.\\

\noindent
The Symanzik polynomials in this given limit are,
\begin{align}
\mathcal{U}=\alpha_1+\alpha_2+\alpha_3\\
\mathcal{F}=x \alpha_3^2+x \alpha_1 \alpha_3+x \alpha_2 \alpha_3+Q^2 \alpha_1 \alpha_2
\end{align}
The top facet scalings, obtained from ASPIRE are $\lbrace -1,-1,-1 \rbrace$ and $\lbrace 0,0,-1 \rbrace$.
\subsubsection{The integral for the scaling $\lbrace-1,-1,-1\rbrace$ }
The Symanzik polynomials for the scaling $\lbrace -1,-1,-1 \rbrace$ in the limit $m^2\gg Q^2$ are
\begin{equation}
\mathcal{U}=\alpha_1+\alpha_2+\alpha_3
\end{equation}
\begin{equation}
\mathcal{F}= x \alpha_3(\alpha_1+\alpha_2+\alpha_3)
\end{equation}
The integral is given by,
\begin{align}\label{tri}
& I^{\lbrace -1,-1,-1 \rbrace}\nonumber \\ \nonumber = & e^{\gamma_E \epsilon}\Gamma(3-\frac{D}{2})\int_{0}^{1}d\alpha_1 \int_{0}^{1} d\alpha_2 \int_{0}^{1} d\alpha_3 \quad \delta(1-\alpha_1-\alpha_2-\alpha_3) \frac{\lbrace x \alpha_3(\alpha_1+\alpha_2+\alpha_3)\rbrace ^{\frac{D}{2}-3}}{(\alpha_1+\alpha_2+\alpha_3)^{D-3}}\nonumber \\ = & e^{\gamma_E \epsilon}\Gamma(3-\frac{D}{2})\quad x^{\frac{D}{2}-3} \int_{0}^{1}d\alpha_1 \int_{0}^{1-\alpha_1} d\alpha_2 (1-\alpha_1-\alpha_2)^{\frac{D}{2}-3}\nonumber \\  = & e^{\gamma_E \epsilon}\Gamma(1+\epsilon)\quad \left(\frac{m^2}{Q^2}\right)^{-1-\epsilon} \left(Q^2 \right)^{-1-\epsilon} \quad \frac{1}{\epsilon (\epsilon-1)} 
\end{align} \\
While looking for the Symanzik polynomials $\mathcal{U,F}$, we consider $m^2\rightarrow x$. But as our expansion parameter is $\frac{m^2}{Q^2}$, in order for writing the result~eq.(\ref{tri}) in terms of expansion parameter with correct consideration, we substitute $x=\frac{m^2}{Q^2}\times Q^2$. After some simplification this is exactly equal to eq.(\ref{actualtop}) for $n=3$. 

\subsubsection{The maximal cut}

\begin{figure}[h]
\centering
  \includegraphics[width=50mm,scale=0.7]{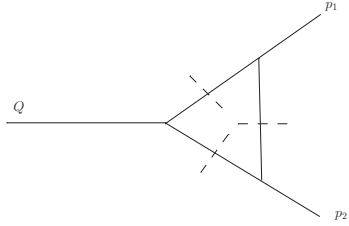}
  \caption{Maximal cut of the triangular diagram.}
  \label{Triangularcut}
\end{figure}
\noindent
In the maximal cut condition, all the propagators are replaced by Dirac Delta function and hence the cut integral is given by
\begin{equation}
I^{MC}_{Triangle}=e^{\gamma_E \epsilon}\int \frac{d^Dk}{i\pi^{\frac{D}{2}}} \delta(k^2-2p_1.k)\delta(k^2-2p_2.k)\delta(k^2-m^2)
\end{equation}
As in the previous case here also we have the angular part trivial to solve with $c=n=3$, $D=4-2\epsilon$ and 
\[ Y_C = \begin{vmatrix}
0 & -\frac{Q^2}{2} & \frac{m^2}{2} \\ -\frac{Q^2}{2} & 0 & \frac{m^2}{2} \\ \frac{m^2}{2} & \frac{m^2}{2} & m^2
\end{vmatrix} = -\frac{m^2Q^2(Q^2+m^2)}{4}\]
and \[ Gram_C = \begin{vmatrix}
0 & -\frac{Q^2}{2} \\ -\frac{Q^2}{2} & 0
\end{vmatrix} = -\frac{(Q^2)^2}{4}\]
Thus using eq.(\ref{cut_master}) we obtain the expression for the maximal cut of this diagram,
\begin{equation}
I^{MC}_{Triangle}=\frac{ e^{\gamma_E \epsilon}}{ 4\Gamma(1-\epsilon)} \frac{1}{\sqrt{m^2 Q^2(m^2+Q^2)}}\left(-\frac{4m^2(Q^2+m^2)}{Q^2}\right)^{-\epsilon}
\end{equation}
Here we have used the following result for the angular integration:
\begin{equation}
    \int d\Omega_{1-2\epsilon} = \frac{2 \pi^{1-\epsilon}}{\Gamma(1-\epsilon)}
\end{equation}

%\subsubsection{The next-to-maximal -cut}
%\begin{equation}
%I^{NMC}_{Triangle}=\int \frac{d^Dk}{i \pi^{D/2}} \frac{\delta(k^2-2p_1.k)\delta(k^2-2p_2.k)}{(k^2-m^2)}
%\end{equation}
%Using eq.(4.41) of \cite{Abreu:2017ptx} the next-to-maximal -cut of the triangular diagram is,
%\begin{equation}
%\begin{split}
%I^{NMC}_{Triangle}=\frac{(2\pi i)2^{-1-2\epsilon} e^{\gamma_E \epsilon}\Gamma(1-\epsilon)}{\pi \Gamma(2-2\epsilon)\sqrt{m^2 Q^2(m^2+Q^2)}} \left(\frac{(m^2)^2}{4Q^2}\right)^{\frac{1-2\epsilon}{2}}\times & \\ {_2F_1}\left(\frac{1}{2},\frac{1-2\epsilon}{2},\frac{3-2\epsilon}{2},-\frac{(Q^2)^2}{16m^2(m^2+Q^2)}\right)
%\end{split}
%\end{equation}

%\begin{figure}[h]
%\centering
 % \includegraphics[width=50mm,scale=0.8]{NMC.eps}
 % \caption{The next-to-maximal  cut of the triangle diagram.}
  %\label{NMC_tri}
%\end{figure}

%\subsection{One loop box diagram}
%We take the massless on-shell box diagram of \cite{Smirnov:2002pj}(see eq.(8.28)),

%\begin{equation}
%I_{Box}=\int d^dk \frac{1}{(k^2+2p_1.k)(k^2-2p_2.k)k^2(k+p_1+p_3)^2}
%\end{equation}

\subsubsection{Correlation between top and cut integrals}
For this diagram, we find between $I^{\lbrace -1,-1,-1 \rbrace}$ and $I^{MC}_{Triangle}$ to be the following,
%\begin{equation}
%I^{\lbrace -1,-1,-1 \rbrace}=\frac{\Gamma(1-\epsilon)\Gamma(1+\epsilon)}{\epsilon (1-\epsilon)}\frac{Q^2}{m}\left(\frac{m^2+Q^2}{Q^2}\right)^{\frac{1}{2}+\epsilon} \quad I^{MC}_{Triangle}
%\end{equation}

\begin{equation}
I^{MC}_{Triangle}= \frac{ \epsilon (1-\epsilon)}{4\Gamma(1-\epsilon)\Gamma(1+\epsilon)}\frac{m}{Q^2}\left(\frac{m^2+Q^2}{Q^2}\right)^{-\frac{1}{2}-\epsilon} \quad I^{\lbrace -1,-1,-1 \rbrace}
\end{equation}
The maximal cut with top condition imposed can be found out using eq.(\ref{eq:top2_integral}) and using eq.(\ref{coruneqmass}) the correlation for this case is given by
\begin{align}
     I^{\{-1,-1,-1\}}=(-1)^{\frac{D-2}{2}}\pi\hspace{.1cm}\text{cosec}\left(\frac{D \pi}{2}\right)I_{Triangle}^{MC}
 \end{align}
 
\subsection{The integral for the top facet $\lbrace 0,0,-1 \rbrace$}
For this diagram, we have the other top facet $\lbrace 0,0,-1\rbrace$. The corresponding Symanzik polynomials are,
\begin{align}
& \mathcal{U}=\alpha_1+\alpha_2 \\
& \mathcal{F}=Q^2 \alpha_1 \alpha_2+m^2 \alpha_1 \alpha_3+m^2\alpha_2 \alpha_3
\end{align}
So we construct the integral,
\begin{align}
I^{\lbrace 0,0,-1 \rbrace}=\int_{0}^{\infty} d\alpha_1 \int_{0}^{\infty} d\alpha_2 \int_{0}^{\infty} d\alpha_3 (\alpha_1+\alpha_2)^{-D/2} e^{-\frac{Q^2 \alpha_1 \alpha_2+m^2 \alpha_1 \alpha_3+m^2\alpha_2 \alpha_3}{\alpha_1+\alpha_2}}
\end{align}
We obtain, 
\begin{align}
I^{\lbrace 0,0,-1 \rbrace}=\frac{\Gamma(-D/2+2)\Gamma^2(D/2-1)}{\Gamma(D-2)}\frac{(Q^2)^{D/2-2}}{m^2}
\end{align}

\subsection{A non-planar two loop diagram}
\begin{figure}[h]
\centering
  \includegraphics[width=60mm,scale=0.9]{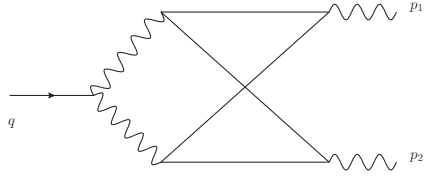}
  \caption{A non-planar two loop diagram.}
  \label{2loop}
\end{figure}

Let us consider a non-planar two loop triangular diagram\ref{2loop}. This diagram had been considered in Ref.\cite{Harley:2017qut,Primo:2016ebd} with $p_1^2=0$, $p_2^2\neq 0$. The integral is defined to be the following,
\begin{align}
I(q^2, m^2,D)= e^{2\gamma_E \epsilon}\int \frac{d^Dk_1}{i\pi^{D/2}} \frac{d^Dk_2}{i\pi^{D/2}} \frac{1}{(k_1-p_1)^2 ((k_2-p_1)^2-m^2) (k_1+p_2)^2((k_1-k_2+p_2)^2-m^2)}  \nonumber \\  \times \frac{1}{((k_1-k_2)^2-m^2)(k_2^2-m^2)}
\end{align}
We consider the limit $p_1^2=0,$ $p_2^2=0,$ $2p_1.p_2=q^2$ and construct the expansion parameter $\frac{m^2}{q^2}$.\\
\\
The Symanzik polynomials are,
\begin{align}
\mathcal{U}=\alpha_1 \alpha_{2456}+\alpha_2 \alpha_{345}+\alpha_3 \alpha_{45}+\alpha_{345} \alpha_6, &\\
\mathcal{F}=x \alpha_{2456} \left(\alpha_3 \alpha_{45}+\alpha_2 \left(\alpha_3+\alpha_{45}\right)+\alpha_6 \left(\alpha_3+\alpha_{45}\right)+\alpha_1 \alpha_{2456}\right) - \nonumber \\ q^2 \left(\alpha_2 \alpha_3 \alpha_5+  \alpha_1 \left(\alpha_6 x_{34}+\alpha_3 \left(\alpha_2+\alpha_{45}\right)\right)\right),
\end{align}
where $\alpha_{ijk\cdots}=\alpha_i+\alpha_j+\alpha_k+\cdots $.\\
The integral has two top facet scalings $\lbrace -1,-1,-1,-1,-1,-1 \rbrace$ and $\lbrace 0,-1,0,-1,-1,-1\rbrace$ for this given limit.
\subsubsection{The integral for the scaling $\lbrace -1,-1,-1,-1,-1,-1 \rbrace$}
The Symanzik polynomials are given by,
\begin{equation}
\mathcal{U}=\alpha_1 \alpha_{2456}+\alpha_2 \alpha_{345}+\alpha_3 \alpha_{45}+\alpha_{345} \alpha_6,
\end{equation}
\begin{equation}
\mathcal{F}=x \alpha_{2456}\left(\alpha_1 \alpha_{2456}+\alpha_2 \alpha_{345}+\alpha_3 \alpha_{456}+\alpha_{45} \alpha_6\right), 
\end{equation}\\
In Ref.\cite{Mishima:2018olh}, four point functions in the high energy limit have been calculated in a systematic way using MoR. While calculating the integrals using MoR, there are regions for which one cannot just use the dimensional regularization, extra analytic regulators \cite{Becher:2011dz,Smirnov:2002pj} are necessary to regularize the contributions from those regions. After obtaining the regions, the parametric integrals have been calculated using the following representation,
\begin{equation}\label{go}
I_{parametric}=\int \mathcal{D}^n\alpha \quad \mathcal{U}^{-D/2}\quad e^{-\frac{\mathcal{F}}{\mathcal{U}}},
\end{equation}\\
where the integral measure is given by,
\begin{equation}
\int \mathcal{D}^n \alpha \equiv \prod_{i=1}^{n}\int_{0}^{\infty}\frac{d\alpha_i  \alpha_i^{\delta_i}}{\Gamma(1+\delta_i)},
\end{equation}
with the consideration of the analytic regulators $\delta_i$. We consider $\delta_i \rightarrow 0 $ while evaluating the parametric integral for the obtained scaling.\\
Thus, we construct the parametric integral for the scaling $\lbrace -1,-1,-1,-1,-1,-1 \rbrace$,
\begin{align}\label{2loopint}
I^{\lbrace -1,-1,-1,-1,-1,-1 \rbrace}&=e^{2\gamma_E \epsilon} \int_{0}^{\infty} \prod_{i=1}^{6} d\alpha_i \quad {\mathcal{U}}^{-\frac{D}{2}} \quad e^{-\frac{\mathcal{F}}{\mathcal{U}}}\nonumber \\
&=e^{2\gamma_E\epsilon}\int_{0}^{\infty}\prod_{i=1}^{6}d\alpha_i(\alpha_{13}\alpha_{2456}+\alpha_{26}\alpha_{45})^{-D/2} e^{-x \alpha_{2456}}
\end{align}
We make the following change of variables,  
\begin{align}
\alpha_1 \rightarrow z_1  z_3, \alpha_2\rightarrow z_2 z_4 z_5, \alpha_3 \rightarrow z_1 (1-z_3), \alpha_4\rightarrow z_2 \left(1-z_4\right) \left(1-z_6\right), \alpha_5 \rightarrow z_2 \left(1-z_4\right) z_6,\nonumber \\ \alpha_6\rightarrow z_2 z_4 \left(1-z_5\right)
\end{align}
The Jacobian of the above transformations is $z_1 z_2^3 \left(1-z_4\right) z_4$. The limits of the new integration variables are the following: 
\[z_1\in [0,\infty],\quad z_2\in [0,\infty],\quad z_3\in [0,1],\quad z_4\in [0,1],\quad z_5\in [0,1],\quad \text{and} \quad z_6\in [0,1]\]
We get,
\begin{align}\label{red_para}
&I^{\lbrace -1,-1,-1,-1,-1,-1 \rbrace}=\nonumber\\
&e^{2\gamma_E \epsilon} \int_{0}^{\infty} dz_1 \int_{0}^{\infty}dz_2\int_{0}^{1}dz_4 z_1 z_2^3 \left(1-z_4\right) z_4 \lbrace z_2 \left(z_2 \left(1-z_4\right) z_4 +z_1\right)\rbrace^{-D/2}e^{-x z_2}
\end{align}
%We now expand the term $ \left(z_2\left(1-z_4\right) z_4 +z_1\right) ^{-D/2}$ binomially to arrive at the following,
%\begin{align}
%I^{\lbrace -1,-1,-1,-1,-1,-1 \rbrace}=\sum_{i=0}^{\infty}\frac{\Gamma(-D/2+1)}{\Gamma(i+1)\Gamma(-D/2-i+1)} \int_{0}^{\infty} dz_1 z_1^{i+1} \int_{0}^{\infty}dz_2 z_2^{3-D-i} e^{x z_2}\times & \\ \nonumber \int_{0}^{1}dz_4 z_4^{-D/2-i+1}\left(1-z_4\right)^{-D/2-i+1} 
%\end{align}

%The $z_1$-integral can be written to be the following, \[\int_{0}^{\infty} dz_1 z_1^{i+1}=\frac{\Gamma(i+2)\Gamma(-i-2)}{\Gamma(0)}\]\\
%We use the reflection formula for gamma function, for integer n, $\frac{\Gamma(z-n)}{\Gamma(z)}=(-1)^n\frac{\Gamma(1-z)}{\Gamma(1-z+n)}$ to write the following (in the limit $z\rightarrow 0$), \[\frac{\Gamma(-i-2)}{\Gamma(0)}=(-1)^{i+2}\frac{1}{\Gamma(3+i)}\]

%Thus, we obtain,
%\begin{align}
%I^{\lbrace -1,-1,-1,-1,-1,-1 \rbrace}=\sum_{i=0}^{\infty}\frac{\Gamma(-D/2+1)}{\Gamma(i+1)\Gamma(-D/2-i+1)}(-1)^{i}\frac{1}{\Gamma(3+i)}\times & \\ \nonumber \frac{\Gamma(4-D-i)\Gamma^2(-D/2-i+2)}{\Gamma(4-D+2i)}x^{D+i-4}
%\end{align}
%We thus obtain, after performing the $z_2$ and $z_4$-integrals, in $D=4-2\epsilon,$

%\begin{align}\label{twolooppara}
%I^{\lbrace -1,-1,-1,-1,-1,-1 \rbrace}= \sum_{i=0}^{\infty} (-1)^i \frac{(i+1)\Gamma(-1+\epsilon)\Gamma(2\epsilon-i)\Gamma(-1+\epsilon-i)}{\Gamma(3+i)\Gamma(-2+2\epsilon-i)} (\frac{m^2}{q^2})^{-2\epsilon+i} (q^2)^{-2 \epsilon+i}
%\end{align}
\noindent We perform the $z_1$-integral with the help of the following formula,
\begin{align}
\int_{0}^{\infty}dz \quad  z^{n_1}(a+z)^{n_2}=\frac{a^{1+n_1+n_2}\Gamma(n_1+1)\Gamma(-1-n_1-n_2)}{\Gamma(-n_2)}
\end{align}
Thus, we obtain,
\begin{align}
I^{\lbrace -1,-1,-1,-1,-1,-1 \rbrace}=\frac{e^{2\gamma_E \epsilon}\Gamma(2)\Gamma(D/2-2)}{\Gamma(D/2)} \int_{0}^{\infty} dz_2 z_2^{5-D} e^{-x z_2}\int_{0}^{1} dz_4 z_4^{3-D/2}(1-z_4)^{3-D/2}
\end{align}
In $D=4-2\epsilon$\footnote{We thank Sumit Banik for the independent check of the analytic expression for this parametric integral using a suitable form of the Method of Brackets.},
\begin{align}\label{twolooptop}
I^{\lbrace -1,-1,-1,-1,-1,-1 \rbrace}=\frac{e^{2\gamma_E \epsilon}\Gamma(-\epsilon)\Gamma(2+2\epsilon)\Gamma^2(2+\epsilon)}{\Gamma(2-\epsilon)\Gamma(4+2\epsilon)} \left(\frac{m^2}{q^2}\right)^{-2-2\epsilon}\left(q^2\right)^{-2-2\epsilon}
\end{align}

\subsection{The maximal cut}
The maximal cut integral is given by
\begin{align}\label{twoloopmaxcut}
I^{MC}_{nonplanar}= e^{2\gamma_E \epsilon}\int \frac{d^Dk_1}{i\pi^{D/2}} \frac{d^Dk_2}{i\pi^{D/2}} \delta(k_1^2) \delta((q-k_1)^2) \delta((p_2-k_2)^2-m^2) \delta((q-k_1-k_2)^2-m^2)\times & \nonumber \\  \delta((k_1+k_2-p_2)^2-m^2) \delta(k_2^2-m^2)
\end{align}
\begin{figure}[h]
\centering
  \includegraphics[width=70mm,scale=0.9]{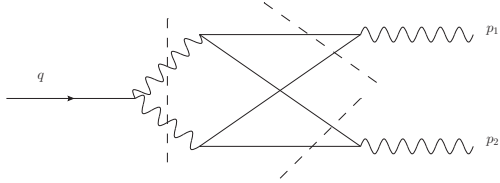}
  \caption{The maximal cut of the non-planar two loop diagram.}
  \label{2loopcut}
\end{figure}
We can evaluate this integral by first using the maximal cut of the box integral involving $k_2$ as the loop variable and then doing remaining integration of $k_1$ variable with the acquired result: 
\begin{equation}  
\begin{split}
    I^{MC}_{nonplanar}= e^{2\gamma_E \epsilon}\int \frac{d^Dk_1}{i\pi^{D/2}}  \delta(k_1^2) \delta((q-k_1)^2) 
    \bigg[\int \frac{d^Dk_2}{i\pi^{D/2}} \delta((p_2-k_2)^2-m^2) \delta((q-k_1-k_2)^2-m^2) & \\ \times  \delta((k_1+k_2-p_2)^2-m^2) \delta(k_2^2-m^2) \bigg]
\end{split}
\end{equation}
We will evaluate the expression inside the square bracket first which is equal to the maximal cut for a massive box diagram using eq.(\ref{cut_master}). This gives,
\begin{equation}\label{2loopmaxcut}
    I^{MC}_{nonplanar}= e^{2\gamma_E \epsilon}\int \frac{d^Dk_1}{i\pi^{D/2}}  \delta(k_1^2) \delta((q-k_1)^2) \frac{\Gamma(1-\epsilon)}{\Gamma(1-2\epsilon)} \frac{1}{\sqrt{Y_C}}\left(\frac{Y_C}{Gram_C}\right)^{-\epsilon},
\end{equation} 
where, \[Gram_C = \begin{vmatrix} \scriptstyle (k_1-q)\cdot(k_1-q) & \scriptstyle (q-k_1)\cdot(p_2-k_1) & \scriptstyle (q-k_1)\cdot(p_2) \\
                                 \scriptstyle (q-k_1)\cdot(p_2-k_1) & \scriptstyle (p_2-k_1)\cdot(p_2-k_1) & \scriptstyle (p_2)\cdot(p_2-k_1) \\
                                 \scriptstyle (q-k_1)\cdot(p_2) & \scriptstyle (p_2)\cdot(p_2-k_1) & \scriptstyle p_2^2 \end{vmatrix} \] 
and   
\begin{equation}
\begin{medsize}
\setlength\arraycolsep{2pt}
Y_C = 
\begin{vmatrix}
\scriptstyle m^2 & \scriptstyle m^2-\frac{1}{2} (k_1-q).(k_1-q) & \scriptstyle m^2-\frac{1}{2} \left(k_1-p_2\right).\left(k_1-p_2\right) & \scriptstyle m^2-\frac{p_2.p_2}{2} \\
 \scriptstyle m^2-\frac{1}{2} (k_1-q).(k_1-q) & \scriptstyle m^2 & \scriptstyle m^2-\frac{p_1.p_1}{2} & \scriptstyle m^2-\frac{1}{2} \left(k_1-p_1\right).\left(k_1-p_1\right) \\
 \scriptstyle m^2-\frac{1}{2} \left(k_1-p_2\right).\left(k_1-p_2\right) & \scriptstyle m^2-\frac{p_1.p_1}{2} & \scriptstyle m^2 & \scriptstyle m^2-\frac{k_1.k_1}{2} \\
 \scriptstyle m^2-\frac{p_2.p_2}{2} & \scriptstyle m^2-\frac{1}{2} \left(k_1-p_1\right).\left(k_1-p_1\right) & \scriptstyle m^2-\frac{k_1.k_1}{2} & m^2 \\
\end{vmatrix}
\end{medsize}
\end{equation}
\\[\baselineskip]
Now to carry out this integral, without loss of generality we can select our frame and parametrize the loop momentum as follows:
\begin{align}\label{eq:parametrization_triangle}
q = \sqrt{q^2}(1,0,{\bf 0}_{D-2}), p_2 = \sqrt{p_2^2}(\alpha,\sqrt{\alpha^2-1},{\bf 0}_{D-2}),
k_1 =  (k_{10},|k_1| \cos \theta,|k_1|\sin\theta ~{\bf 1}_{D-2})
\end{align}
where $\theta \in [0,\pi]$ and $|k|>0$, and ${\bf 1}_{D-2}$ ranges over unit vectors in the dimensions transverse to $q$ and $p_2$.
Momentum conservation fixes the value of $\alpha$ in terms of the momentum invariants to be 
\begin{equation}
\alpha = \frac{q^2-p_2^2}{2 \sqrt{q^2}\sqrt{p_2^2}}
.
\end{equation}
In $D$ dimensions, we have
\begin{equation}
    d^Dk_1 = dk_{10}\,|k_1|^{D-2}\,d|k_1|\,d\phi \,\sin \theta_1 \,d\theta_1\, \sin^2 \theta_2 \,d\theta_2 \,...\,\sin^{D-3}\theta_{D-3}\,d\theta_{D-3}
\end{equation}
Thus in $D=4-2\epsilon$ dimensions after doing the $\phi$ integration in the remaining $D-2$ dimensions we get
\begin{equation}\label{kMeasure}
    d^{D} k~\delta(k_1^2) = d k_{10} ~d|k_1|~d\cos\theta~\delta(k_{10}^2-|k_1|^2)~
  \frac{2 \pi^{1-\epsilon}}{\Gamma(1-\epsilon)} |k_1|^{2-2\epsilon} (\sin\theta)^{-2\epsilon}\,.
\end{equation}
Thus eq.(\ref{2loopmaxcut}) becomes:
\begin{equation}
\begin{split}
    I^{MC}_{nonplanar}=\frac{2e^{2\gamma_E \epsilon}}{i\pi}\int_{0}^{\infty} 
d k_{10} ~ \int_0^\infty d|k_1|~ \int_{-1}^{1} d\cos\theta~\delta(k_{10}^2-|k_1|^2)~
  \frac{ |k_1|^{2-2\epsilon} (\sin\theta)^{-2\epsilon}}{\Gamma(1-2\epsilon)}  & \\
  \times \frac{\delta(q^2-2k_{10}\sqrt{q^2}+k_{10}^2-|k_1|^2)}{\sqrt{Y_C}}\left(\frac{Y_C}{Gram_C}\right)^{-\epsilon}
  \end{split}
\end{equation}
Now the integrations in $k_{10}$ and $|k_1|$ are trivial owing to the existence of the delta functions \footnote{Here we have enforced the condition $k_{10}>0$ for evaluating the delta function (see for reference eq.(3.6) in \cite{Abreu:2014cla})}. Thus after performing these integrations and enforcing the condition $p_1^2=p_2^2=0$, we get,
\begin{equation}
    I^{MC}_{nonplanar}=\frac{2\pi e^{2\gamma\epsilon}}{\Gamma(1-2\epsilon)}(\sqrt{q^2})^{-3-2\epsilon}\int_{-1}^{1}d\cos\theta\left(m^2+\frac{q^2}{16}(1-\cos^2\theta)\right)^{-\frac{1}{2}-\epsilon}\left(1-\cos^2\theta\right)^{-\frac{1}{2}-\epsilon}
\end{equation}
Performing the change of variables
%\begin{equation}
%\cos \theta = 2 u - 1, \qquad u \in [0,1],
%\end{equation}
\begin{equation}
\cos \theta = u
\end{equation}
we get,
\begin{align}
 I^{MC}_{nonplanar}=\frac{4\pi e^{2\gamma\epsilon}}{\Gamma(1-2\epsilon)}(\sqrt{q^2})^{-3-2\epsilon}\int_{0}^{1}du\left(m^2+\frac{q^2}{16}(1-u^2)\right)^{-\frac{1}{2}-\epsilon}\left(1-u^2\right)^{-\frac{1}{2}-\epsilon}
\end{align}
We finally obtain,
\begin{align}
 I^{MC}_{nonplanar}=\frac{4\pi e^{2\gamma\epsilon}}{\Gamma(1-2\epsilon)}(\sqrt{q^2})^{-3-2\epsilon}\frac{\Gamma(1/2)\Gamma(1/2-\epsilon)}{\Gamma(1-\epsilon)}(m^2+\frac{q^2}{16})^{-\frac{1}{2}-\epsilon} \times & \\ \nonumber {}_2F_{1}(1/2+\epsilon,1/2;1-\epsilon;\frac{q^2}{q^2+16m^2})
\end{align} 
%\begin{equation}
   % I^{MC}_{nonplanar}=\frac{2\pi e^{\gamma\epsilon}}{\Gamma(1-2\epsilon)}(\sqrt{q^2})^{-3-2\epsilon}\int_{0}^{1}du\left(m^2+\frac{q^2}{4}u(1-u)\right)^{-\frac{1}{2}-\epsilon}\left(u(1-u)\right)^{-\frac{1}{2}-\epsilon}
%\end{equation}
%The first bracket inside the integral can be expanded in a binomial expansion,
%\begin{align}
   % I^{MC}_{nonplanar}=\frac{2\pi e^{\gamma\epsilon}}{\Gamma(1-2\epsilon)}(\sqrt{q^2})^{-3-2\epsilon}\sum_{i=0}^{\infty}\frac{\Gamma(-1/2-\epsilon+1)}{\Gamma(-1/2-\epsilon-i+1)\Gamma(i+1)} (m^2)^{-\frac{1}{2}-\epsilon-i}\left(\frac{q^2}{4}\right)^i \times & \\ \nonumber \int_{0}^{1}du\left(u(1-u)\right)^{-\frac{1}{2}-\epsilon+i}
%\end{align}
%This can be evaluated using Gamma functions and finally we get,
%\begin{align}\label{mc_nonplanar}
%I^{MC}_{nonplanar}=\frac{2\pi e^{\gamma\epsilon}}{\Gamma(1-2\epsilon)}(\sqrt{q^2})^{-3-2\epsilon}(m^2)^{-\frac{1}{2}-\epsilon}\sum_{i=0}^{\infty}\frac{\Gamma(-1/2-\epsilon+1)}{\Gamma(-1/2-\epsilon-i+1)\Gamma(i+1)} \times  & \\ \nonumber \frac{\Gamma^2(-\frac{1}{2}-\epsilon+i)}{\Gamma(-1-2\epsilon+2i)} \left(\frac{q^2}{4m^2}\right)^i
%\end{align}

\subsection{Correlation between $I^{\lbrace -1,-1,-1,-1,-1,-1\rbrace}$ and the maximal  cut}
For the non-planar diagram~\ref{2loop} in the given limit, we obtain the following relation :
\begin{align}
I^{MC}_{non-planar}= \frac{4\pi\Gamma(1/2)\Gamma(1/2-\epsilon)\Gamma(2-\epsilon)\Gamma(4+2\epsilon)}{\Gamma(1-\epsilon)\Gamma(1-2\epsilon)\Gamma(-\epsilon)\Gamma(2+2\epsilon)\Gamma^2(2+\epsilon)}\left(\frac{m^2}{q^2}\right)^{\frac{3}{2}+\epsilon}\times & \\ \nonumber \left(1+\frac{q^2}{16m^2}\right)^{-\frac{1}{2}-\epsilon} {}_2F_{1}\left(1/2+\epsilon,1/2;1-\epsilon;\frac{q^2}{q^2+16m^2}\right)\quad I^{\lbrace -1,-1,-1,-1,-1,-1\rbrace}
\end{align}
Once again we will find out the maximal cut with top condition imposed. Here since we do not have the general formulation of the maximal cut for more than one loop we will evaluate the maximal cut with zero external momenta using the method of residues particularly for this case. Using eq.(\ref{twoloopmaxcut}) the maximal cut integral with top condition imposed is given by

\begin{equation}  
\begin{split}
    & I^{MC}_{nonplanar}= \\ & e^{2\gamma_E \epsilon}
    \int \frac{d^Dk_1}{i\pi^{D/2}}  \delta(k_1^2) \delta(k_1^2) 
    \bigg[\int \frac{d^Dk_2}{i\pi^{D/2}} \delta(k_2^2-m^2) \delta((k_1+k_2)^2-m^2)  \delta((k_1+k_2)^2-m^2) \delta(k_2^2-m^2) \bigg]
\end{split}
\end{equation}
We will do the inner one loop integral using the method of residues\cite{Abreu:2017ptx}. The integration variable here is $k_2$ and $k_1$ can be considered as a constant for the inner integration. Here we can see that the four propagators are of two kinds and hence using the method of residues the integral becomes
\begin{align}  
 & I^{MC}_{nonplanar}=\frac{2^{(4D-11)}e^{2\gamma_{E}\epsilon}}{\pi^{\frac{3}{2}}\Gamma\left(\frac{D-3}{2}\right)}\int \frac{d^Dk_1}{i\pi^{D/2}}  \delta(k_1^2) \delta(k_1^2) \\  &
 \times 
\text{Res}_{r^2=-m^2}\left[\text{Res}_{t_0=T_0}\left[\frac{\left(r^2\right)^{\frac{D-2}{2}}}{(r^2+m^2)^2} \frac{\left[t_0(1-t_0)\right]^{\frac{D-3}{2}}}{B_0^2\left( t_0-T_{0}\right)^2}\right]\right]
\prod_{j=1}^{2}\int_0^1 dt_j \left[t_j(1-t_j)\right]^{\frac{D-3-j}{2}}
\end{align}
The innermost integration is again trivial using beta function. Now the pole at $t_0=T_0$ is of order two. Hence using Cauchy's theorem of residues for higher order poles, after eliminating the denominator we have to differentiate the quantity inside once with respect to $t_0$ and then take the limit $t_0 \to T_0$. Doing all this we get
\begin{align}  
 & I^{MC}_{nonplanar}= \nonumber \\
 & \frac{2^{(2D-5)}(D-3)e^{2\gamma_{E}\epsilon}}{\pi^{\frac{1}{2}}\Gamma\left(\frac{D-1}{2}\right)}\int \frac{d^Dk_1}{i\pi^{D/2}}  \delta(k_1^2) \delta(k_1^2)\text{Res}_{r^2=-m^2}\left[\frac{(1-2T_0)\left(r^2\right)^{\frac{D-2}{2}}\left[T_0(1-T_0)\right]^{\frac{D-5}{2}}}{B_0^2\hspace{.1cm}(r^2+m^2)^2} \right]
\end{align}
 Now using section (4.1) of \cite{Abreu:2017ptx} we have
 \begin{align}
     B_0=4 r k_1 \hspace{.5cm}\text{and}\hspace{.5cm}\, T_0=\frac{2 r k_1-m^2-k_1^2-r^2}{4 r k_1}
 \end{align}
 Again the pole at $r^2=-m^2$ is of order two and hence we will do the same process again to find out the residue which gives
 \begin{align}  
  I^{MC}_{nonplanar}=  \frac{(-1)^\frac{D-1}{2}(D-3)e^{2\gamma_{E}\epsilon}}{\pi^{\frac{1}{2}}\Gamma\left(\frac{D-1}{2}\right)}\int \frac{d^Dk_1}{i\pi^{D/2}}  \delta(k_1^2) \delta(k_1^2)\frac{  \left(k_1^2+4m^2\right)^{\frac{D-1}{2}} \left(4 m^2-(D-6) k_1^2\right)}{k_1^3\left(k_1^2+4  m^2\right)^3}
\end{align}
Here we have done the differentiation with respect to $r$ rather than $r^2$ and then divided by $2r$ after differentiation as both the processes yield the same result but its easier to evaluate using the previous method. Once again we are going to use parametrization used in \cite{Abreu:2017ptx} and the method of residues to find out the above integral with residue at $k_1^2=0$, 
 \begin{align}  
  I^{MC}_{nonplanar}=  (-1)^\frac{D-1}{2}\frac{2^{D-2}(D-3)e^{2\gamma_{E}\epsilon}}{\pi\Gamma\left(D-1\right)}\text{Res}_{k_1^2=0}\left[\frac{  \left(k_1^2+4m^2\right)^{\frac{D-1}{2}} \left(4 m^2-(D-6) k_1^2\right)}{(k_1^2)^{\frac{9-D}{2}}\left(k_1^2+4  m^2\right)^3}\right]
\end{align}
So we see that the pole here is of order $\frac{9-D}{2}$ and hence after eliminating the term $(k_1^2)^{\frac{9-D}{2}}$ we have to take the derivative of order $\frac{7-D}{2}$ with respect to $k_1^2$. Doing this we get, 
\begin{align}  
  I^{MC}_{nonplanar}=
  (-1)^{D+1}\frac{2^{3D-14}\Gamma(7-D)(D-3)e^{2\gamma_{E}\epsilon}}{\pi\Gamma\left(\frac{5-D}{2}\right)\Gamma\left(\frac{9-D}{2}\right)\Gamma\left(D-1\right)}(m^2)^{-6+D}
\end{align} 
Using $D=4-2\epsilon$ and after some simplification this becomes, 
\begin{align}  
  I^{MC}_{nonplanar}=
  (-1)^{5-2\epsilon}\frac{2^{2-2\epsilon}\Gamma(3+2\epsilon)\Gamma(3+\epsilon)\Gamma(1+\epsilon)(1-2\epsilon)e^{2\gamma_{E}\epsilon}}{\pi^3\Gamma\left(1-2\epsilon\right)\Gamma\left(5-2\epsilon\right)\Gamma\left(3-2\epsilon\right)}(m^2)^{-2-2\epsilon}
\end{align} 
Again comparing with eq.(\ref{twolooptop}), we find that there is a correlation in the form of proportionality between the two integrals as the power of $m^2$ is equal in both cases and hence after some simplification the correlation becomes,
\begin{align}\label{cortwoloop}
I^{\lbrace -1,-1,-1,-1,-1,-1 \rbrace}=(-1)^{5-2\epsilon}\frac{2^{2\epsilon} \Gamma (1-2 \epsilon ) \Gamma (5-2 \epsilon ) \Gamma (-2 \epsilon )}{\Gamma (5+2 \epsilon )}I^{MC}_{nonplanar}
\end{align}

\subsection{The integral for the top facet $\lbrace 0,-1,0,-1,-1,-1 \rbrace$}
In this case, we compute the integral for the other top facet $\lbrace 0,-1,0,-1,-1,-1\rbrace$. The Symanzik polynomials are,
\begin{align}
& \mathcal{U}=\alpha_{13} \alpha_{2456}\\
& \mathcal{F}=\alpha_{2456}(-q^2 \alpha_1 \alpha_3+m^2 \alpha_{13}\alpha_{2456})
\end{align}
The integral is given by,
\begin{align}
I^{\lbrace 0,-1,0,-1,-1,-1\rbrace}=\frac{\Gamma(-D/2+2)\Gamma^2(D/2-1)\Gamma(-D/2+4)}{6 \Gamma(D-2)}\frac{(-q^2)^{D/2-2}}{(m^2)^{-D/2+4}}
\end{align}
\section{Discussion and conclusion}\label{sec5}
We have considered given multi-scale Feynman diagrams in a given limit and obtained the scalings required for the asymptotic expansion of the diagram. The exploration here, which is based on Landau equations, allows us going beyond the bottom facet results. Furthermore, ASY and ASPIRE were concerned with unveiling the regions. Here Landau equations permit us to explore the consequences of the asymptotic analysis of the Feynman graphs combined with the corresponding maximal cuts.\\  
%In this work, we have explored the parametric integrals constructed from the top facet scaling(with equal components) of the mathematica program ASPIRE. The Landau criteria for a generic Feynman integral give the location of the singularities. We have used this fact to correlate the obtained parametric integrals to the maximal  cut of the corresponding Feynman diagrams.\\
We have brought two independent approaches together, and the existence of the top facets is important to study as it is closely related to the bottom facets on which the MoR rests.  Thus it is imperative to study the consistency of the theory, which required us to study the cut technology.\\ 
A two point one loop diagram, a one loop triangular diagram, and a two loop non-planar diagram have been studied. For these examples, we have found that the integral constructed based on the top facet scaling (with equal components) of the Newton polytope has the correspondence with the maximal cut of the corresponding Feynman diagrams, having the following form in $D=4-2\epsilon$,
\begin{equation}
I^{MC}= f(m_i^2,Q_j^2,\epsilon)\quad  I^{\lbrace -1,-1,-1,\cdots,-1 \rbrace},
\end{equation}
where $m_i$ are the masses of the internal lines, and $Q_j$ are the external momenta.
\\
For the one loop cases, we have derived a generalized formula for the top facet $I^{\lbrace -1,-1,-1,\cdots,-1 \rbrace}$ and find a generalized expression for the correlation factors using the formalism of \cite{Abreu:2017ptx}. For going beyond one loop, it might be helpful to consider the generalization of \cite{Abreu:2017ptx} up to higher loop orders.\\
The top facet having equal components essentially corresponds to the limit where $q^2$ can be neglected with respect to $m^2$, and with this criterion imposed on the maximal cut of the given Feynman integrals, we find that both (the maximal cut and the top facet with equal components) give rise to the same power of $m^2$ (eq.(\ref{coruneqmass}), eq.(\ref{coreqmass}) and eq.(\ref{cortwoloop})), also the same form of the result for the unequal masses case in one loop (eq.(\ref{coruneqmass})). Thus we find that the maximal cut in the asymptotic limit is exactly proportional to the top integrals having equal components for all of the studied examples.
\begin{equation}
I^{MC}= f(\epsilon)\quad  I^{\lbrace -1,-1,-1,\cdots,-1 \rbrace},
\end{equation}
where $m_i$ are the masses of the internal lines and $Q_j$ are the external momenta.\\
The observed proportionality indicates a deep sense of correlation between top integrals and maximal cut.  From the Landau equations, the on-shell condition of all propagators, i.e., the maximal cut condition, is possible only when all $\alpha_i's\neq0$. Since in the integration domain of $\alpha_i's$ this corresponds to all the points in the domain except the point where all $\alpha_i's=0$ that is the origin, for the top integral whose integration domain includes all the points including the origin, the result is not much different from that of the maximal cut except the prefactors which are functions of $\epsilon$. This proportionality of the result and similarity in the integration domain is the correlation between the top integral and maximal cut.\\
The prediction of correlation can also be obtained from the fact that for a meromorphic function (the integrand of the Feynman integral in our case), the integral over it for a single variable in a complex plane is related to the residues of the singularities inside the contour integral \cite{arfken}. Though this reasoning is not mathematically rigorous but definitely throws some light on the reason behind the correlation in the form of proportionality.\\
We have also computed the other top facet integrals obtained in the examples we have considered. For this kind of top integrals, a loop momentum representation is not easily expressible as some $\alpha_i's$ are non leading in comparison to others. That's why we have found out an alpha parametric representation for cuts for the one loop case (eq.(\ref{alphacut34})) in order to correlate with those other scalings in alpha representation in appendix \ref{cutalpha} if possible in a future research project. \\ 
The general formula of cuts for more than one loop is yet to be found in the literature. So for more than one loop case, we have to specifically evaluate the maximal cut and the top integral for each case and find out whether there is a correlation. We have worked out a two loop example and found out that the kind of proportionality that we have for the one loop cases also exists for this particular two loop example (eq.(\ref{cortwoloop})). We are not sure that this is true for every other many loop example, for that we need a generalization of cuts to the many loop cases, which is a future research project.\\
The limit of expansion parameter tending to infinity is equivalent to the asymptotic expansion of a given Feynman diagram in the large mass expansion. There are prescriptions in the literature to deal with the large mass expansion in the language of expansion by sub-graphs\cite{Smirnov:2002pj,Smirnov:1994tg,Fleischer:1998nc,Fleischer:1998cr}. Implementation of such a prescription in the framework ASPIRE is a topic of future investigation. \\ 
In this work, we have taken up the subject of the top facets that have arisen in the ASPIRE algorithm (note that the ASY algorithm can also be used to generate ) as a result of asymptotic analysis and Landau equations.  In order to test the consistency of the asymptotic expansion, which in the context of the bottom facets leads to MoR, we have carried out a detailed study and have linked it to the hitherto unrelated topic of cut Feynman diagrams.  In the previous sections, we have given a thorough exposition of all the aspects of our study, pointing out the strengths as well as topics to be studied in the future.  This is a novel approach that has been used to study features of one-loop integrals in their entirety as well as a non-trivial two-loop integral. This would be the first of what could be a series of explorations using these seemingly unrelated methods. Also, it is conceivable that one could relate results from the asymptotic analysis to those coming from the studies of Hopf Algebras \cite{Abreu:2014cla} as in the case of Feynman integrals with multiple polylogarithms and via dispersion relations.
 
\section*{Acknowledgements}
We are grateful to V.A. Smirnov and Go Mishima for carefully reading the manuscript and a lot of valuable comments. It is a pleasure to thank Abhishek Pal, Samuel Friot, Sumit Banik, Daniel Wyler and Amit Adhikary for many useful discussions during the course of this work. We thank Go Mishima for sharing useful calculations on the evaluation of parametric integral. RS likes to acknowledge the support of INSPIRE Fellowship[IF150859]. JaxoDraw\cite{Binosi:2008ig} has been used for drawing all the Feynman diagrams considered in this work.

\appendix
\section{Appendix}\label{appenA}

\subsection{Brief description of the ancillary files}

\begin{table}[h]
    \begin{tabular}{|c|c|}
       \hline
File name & Description \\
\hline
OneloopVertex.nb & A one loop vertex integral has been analyzed.\\
\hline
TwoPointOneLoop.nb & A two point one loop diagram \\ & has been analyzed. \\
\hline
ScalarTriangle.nb & A scalar triangular diagram \\ & has been analyzed.\\
\hline 
TwoLoopNonPlanar.nb & A non-planar two loop triangle diagram has been analyzed. \\
\hline
    \end{tabular}
    \caption{Description of Mathematica notebooks used in this work.}
    \label{nb}
\end{table}

\subsection{Comparison of the scales obtained using ASPIRE and ASY}\label{topBottom}
In this section, we summarize the technical aspects of our consideration for the bottom and top facets of the Newton polytope obtained from the sum of the Symanzik polynomials with suitable linear tranformations.\\
The bottom facets are those facets of the Newton polytope for which 
\[\vec{r}.\vec{v}=c, \quad \text{for the points $\vec{r}$ lying on the facets,}\]
\[\vec{r}.\vec{v}> c, \quad \text{for the points $\vec{r}$ lying above the facets,}\]
where $\vec{r}$ are the vector exponents of the terms of a given sum for the construction of the Newton polytope and $\vec{v}$ are the normal vectors corresponding to the facets of the Newton polytope.\\
For bottom facets, we consider the limit $x=\frac{m^2}{q^2}\rightarrow 0 $ $(\text{i.e.}\quad m^2\ll q^2)$ and $q^2\rightarrow 1 $. This is the well-known case of ``Regions".\\ 
The top facets are those facets of the Newton polytope for which
\[\vec{r}.\vec{v}=c, \quad \text{for the points $\vec{r}$ lying on the facets.}\]
\[\vec{r}.\vec{v}< c, \quad \text{for the points $\vec{r}$ lying below the facets.}\]
For the case of top facets, we utilize the freedom of considering the other possibility to take the expansion parameter $x=\frac{m^2}{q^2}\rightarrow \infty$ $(\text{i.e.}\quad m^2\gg q^2)$ and we do not impose the constraint $q^2\rightarrow 1$ while computing the Symanzik polynomials. This corresponds to the expansion of the Feynman graphs in the large mass limit.\\
It is trivial to see that the limit $x=\frac{m^2}{q^2}\rightarrow 0$ is equivalent to the limit $x=\frac{q^2}{m^2}\rightarrow \infty$ and vice versa. This implies one can transform the bottom facets into top facets with the transformed limits and vice versa.\\
We here present in the table\ref{comp} the explicit comparison of the scaling coming from the bottom and top facets using ASPIRE and ASY for the given examples.
\begin{table}[htb!]
\begin{bigcenter}
\scalebox{0.95}{
\begin{tabular}{|c|c|c|c|c|}
\hline
\multirow{3}{*}{Diagrams} & \multicolumn{2}{c|}{ASPIRE} & %
    \multicolumn{2}{c|}{ASY} \\
\cline{2-5}
 & \multicolumn{2}{c|}{Scaling from} & \multicolumn{2}{c|}{Scaling from} \\
\cline{2-5}
 & \makecell{Bottom facet} & \makecell{top facet} & \makecell{Bottom face} & \makecell{top facet} \\
\hline
 Two point one loop & \makecell{ \{ 0,0 \},\\ \{ -1/2,-1 \},\\ \{-1,-1/2\}} & \{ -1,-1 \}&\makecell{\{0,0\},\\ \{0,1/2\},\\\{0,-1/2\}} &\{0,0\} \\
\hline
 One loop triangle &\makecell{\{ 0,0,0 \},\\ \{ -1,0,-1\},\\ \{ 0,-1,-1 \}} & \makecell{\{ -1,-1,-1 \},\\\{ 0,0,-1 \}}&\makecell{\{0,-1,-1\},\\\{0,0,0\},\\\{0,1,0\}} & \makecell{\{0,0,0\},\\ \{0,0,-1\}}\\
\hline
Two loop non-planar & \makecell{\{0,-1,0,0,0,-1\},\\ \{-1,-1,0,-1,-1,0\},\\ \{-1,-1,0,0,-1,-1\},\\ \{0,0,0,0,0,0\},\\ \{0,-1,-1,-1,0,-1\},\\ \{0,0,0,-1,-1,0\},\\ \{0,-1,0,-1,-1,-1\},\\ \{0,0,-1,-1,-1,-1\}} &\makecell{\{0,-1,0,-1,-1,-1,\},\\ \{-1,-1,-1,-1,-1,-1\}} & \makecell{ \{0,-1,-1,-1,0,-1\},\\ \{0,-1,0,-1,-1,-1\},\\ \{0,-1,0,0,0,-1\},\\ \{0,0,-1,-1,-1,-1\},\\ \{0,0,0,-1,-1,0\},\\ \{0,0,0,0,0,0\},\\\{0,0,1,0,0,1\},\\ \{0,0,1,1,0,0\}} & \makecell{\{0,0,0,0,0,0\},\\\{0,-1,0,-1,-1,-1\}}\\
\hline
% etc. ...
\end{tabular} }
\end{bigcenter}
\caption{\label{comp}Comparison between ASPIRE and ASY for the given examples.}
\end{table}\\ 
It immediately turns out that the scalings for the bottom and top facets as obtained from the package ASY and ASPIRE match exactly for the given examples.

\subsection{Hypergeometric Function $_{2}F_1(a,b;c;x)$}
The hypergeometric function $_{2}F_1(a,b;c;x)$ is given by,
\begin{equation}
_{2}F_1(a,b;c;x)=\sum_{n=0}^\infty \frac{(a)_n (b)_n}{(c)_n}\frac{x^n}{n!},
\end{equation}
where $(a)_n=\frac{\Gamma(a+n)}{\Gamma(a)}$ is the Pochhammer symbol.
In the integral representation, 
\begin{equation}
_{2}F_1(a,b;c;x)=\frac{\Gamma(c)}{\Gamma(b)\Gamma(c-b)}\int_{0}^{1} u^{b-1}(1-u)^{c-b-1}(1-x u)^{-a}du,
\end{equation}
where $Re(b), Re(c)>0$.
%\subsection{Duplication formula for Gamma function}
%\begin{equation}\label{dup}
  %  \Gamma(2n)=\frac{1}{\sqrt{\pi}}2^{2n-1}\Gamma(n)\Gamma(n+\frac{1}{2})
%\end{equation}
\subsection{Angular Integration}
According to the convention followed in eq.(A.1) in  \cite{Abreu:2017ptx}  which states that 
\begin{equation}
    d^Dk^E=d^{c-1}k_{\parallel}d^{D-c+1}k_{\perp}=\frac{1}{2}d^{c-1}k_{\parallel}d\Omega_{D-c}(k_{\perp}^2)^{(D-c+1)/2}dk_{\perp}^2
\end{equation}
where $k_{\parallel}$ and $k_{\perp}$ are the parallel and perpendicular components to the set of cut propagators, the angular part of the integration is given by
\begin{equation}\label{angularint}
    \int d\Omega_{D} = \frac{2 \pi^{(D+1)/2}}{\Gamma((D+1)/2)}
\end{equation} 
instead of the conventional
\begin{equation}
    \int d\Omega_{D} = \frac{2 \pi^{D/2}}{\Gamma(D/2)}
\end{equation}

\section{Cuts in alpha parametrization for the one loop case}\label{cutalpha}
In this section we write the cuts in alpha parametrization in case of one loop. Consider the following expression in alpha parametrization
\begin{align}\label{alphacut}
    \frac{1}{A B}=\int_0^1\frac{d\alpha}{(\alpha A+(1-\alpha)B)^2}
\end{align} 
If we put $A=1$ this becomes
\begin{align}\label{alphacut1}
    \frac{1}{B}=\int_0^1\frac{d\alpha}{(\alpha +(1-\alpha)B)^2}
\end{align} 
Now consider an integral 
\begin{align}\label{alphacut2}
    \int \frac{dA}{A B}
\end{align}
Lets say we have the on-shell condition $A=0$ like that of a cut integral then this integral becomes
\begin{align}\label{alphacut3}
    2\pi i\int \delta(A) \frac{dA}{B}= \frac{2\pi i}{B_A}
\end{align} 
where $B_A$ is $B$ evaluated when $A=0$. Also this is exactly equal to the case when we evaluate the residues at singularity $A=0$.
\begin{align}\label{alphacut4}
    \oint_{C_A}\frac{dA}{AB}=\frac{2\pi i}{B_A}
\end{align}
where $C_A$ is the contour integral encircling the singularity at $A$. Now using eq.(\ref{alphacut}) to eq.(\ref{alphacut4}) we get
\begin{align}\label{alphacut5}
    2\pi i\int \delta(A) \frac{dA}{B}=\oint_{C_A}\frac{dA}{AB}=\oint_{C_A}dA\int_0^1\frac{d\alpha}{(\alpha A+(1-\alpha)B)^2}=\frac{2\pi i}{B_A}=2\pi i\int_0^1\frac{d\alpha}{(\alpha +(1-\alpha)B_A)^2}
\end{align}
Thus we have
\begin{align}\label{alphacut6}
\oint_{C_A}dA\int_0^1\frac{d\alpha}{(\alpha A+(1-\alpha)B)^2}=2\pi i\int_0^1\frac{d\alpha}{(\alpha +(1-\alpha)B_A)^2}
\end{align} 
Thus we see the effect of evaluating the residue at $A=0$ or finding out the cut integral with cut at $A$ is to set $A=1$ and replacing the other propagators with there values evaluated at $A=0$ in alpha parametrization. Now consider the one loop integral eq.(\ref{eq:uncut_integral})
\begin{align}\label{alphacut7}
I_n=&(-1)^n\frac{2^{\sum_{j=0}^{n-2}(D-2-j)}e^{\gamma_{E}\epsilon}}{\pi^{\frac{n-1}{2}}\Gamma\left(\frac{D-n+1}{2}\right)}\int_0^\infty dr^2\frac{\left(r^2\right)^{\frac{D-2}{2}}}{r^2+m_{n-1}^2}
\prod_{j=0}^{n-2}\int_0^1 dt_j \frac{\left[t_j(1-t_j)\right]^{\frac{D-3-j}{2}}}{B_j\left( t_j-T_{j}\right)}\, ,
\end{align}
Applying Feynman parametrization we get
\begin{align}\label{alphacut8}
I_n=&(-1)^n\frac{2^{\sum_{j=0}^{n-2}(D-2-j)}e^{\gamma_{E}\epsilon}}{\pi^{\frac{n-1}{2}}\Gamma\left(\frac{D-n+1}{2}\right)}\int_0^\infty dr^2\left(r^2\right)^{\frac{D-2}{2}}
\prod_{j=0}^{n-2}\int_0^1 dt_j \left[t_j(1-t_j)\right]^{\frac{D-3-j}{2}}\nonumber \\ 
&\times \prod_{k=0}^{n-1}\int_0^1d\alpha_k\frac{\delta(1-\sum_{l=0}^{n-1}\alpha_l)}{\left(\alpha_{n-1}(r^2+m_{n-1}^2)+\sum_{j1=0}^{n-2}\alpha_{j1}B_{j1}(t_{j1}-T_{j1})\right)^n} ,
\end{align}
Now using eq.(\ref{alphacut6}) we get the cut integral with the method of residues
\begin{align}\label{alphacut9}
&C_cI_n= \nonumber \\
&(-1)^n\frac{2^{\sum_{j=0}^{n-2}(D-2-j)}e^{\gamma_{E}\epsilon}}{\pi^{\frac{n-1}{2}}\Gamma\left(\frac{D-n+1}{2}\right)}
\prod_{j=0}^{c-1} \left[t_{j,p}(1-t_{j,p})\right]^{\frac{D-3-j}{2}} \int_0^\infty dr^2\left(r^2\right)^{\frac{D-2}{2}}\prod_{j=c}^{n-2}\int_0^1 dt_j \left[t_j(1-t_j)\right]^{\frac{D-3-j}{2}}\nonumber \\ 
&\times\prod_{k=0}^{n-1}\int_0^1d\alpha_k\frac{\delta(1-\sum_{l=0}^{n-1}\alpha_l)}{\left(\alpha_{n-1}(r^2+m_{n-1}^2)+\sum_{j2=0}^{c-1}\alpha_{j2}[B_{j2}]_c+\sum_{j1=c}^{n-2}\alpha_{j1}[B_{j1}]_c\left(t_{j1}-[T_{j1}]_c\right)\right)^n} ,
\end{align}
Here $t_{j,p},[B_j]_c$ are the corresponding values of $t_j,B_j$ in the locus of cut respectively and c is equal to the number of cut propagators. Now changing to Schwinger parametrization we get
\begin{align}\label{alphacut10}
&C_cI_n=&\nonumber\\&(-1)^n\frac{2^{\sum_{j=0}^{n-2}(D-2-j)}e^{\gamma_{E}\epsilon}}{\pi^{\frac{n-1}{2}}\Gamma\left(\frac{D-n+1}{2}\right)\Gamma(n)}\prod_{j=0}^{c-1} \left[t_{j,p}(1-t_{j,p})\right]^{\frac{D-3-j}{2}}\int_0^\infty dr^2\left(r^2\right)^{\frac{D-2}{2}}\prod_{j=c}^{n-2}\int_0^1 dt_j \left[t_j(1-t_j)\right]^{\frac{D-3-j}{2}} 
\nonumber\\
&\times \prod_{k=0}^{n-1}\int_0^\infty ds_{k}\hspace{.1cm} exp\left[-\left(s_{n-1}(r^2+m_{n-1}^2)+\sum_{j2=0}^{c-1}s_{j2}[B_{j2}]_c+\sum_{j1=c}^{n-2}s_{j1}[B_{j1}]_c\left(t_{j1}-[T_{j1}]_c\right)\right)\right],
\end{align}
To get the alpha parametrization form we need to integrate out the loop momentum variables. Thus re-expressing the integral so as to do the integration in r and $t_j$ variables first we get
\begin{align}\label{alphacut11}
&C_cI_n=(-1)^n\frac{2^{\sum_{j=0}^{n-2}(D-2-j)}e^{\gamma_{E}\epsilon}}{\pi^{\frac{n-1}{2}}\Gamma\left(\frac{D-n+1}{2}\right)\Gamma(n)}\prod_{j=0}^{c-1} \left[t_{j,p}(1-t_{j,p})\right]^{\frac{D-3-j}{2}}
\nonumber\\
&\times \prod_{k=0}^{n-1}\int_0^\infty ds_{k}\hspace{.1cm} exp\left[-\left(\sum_{j2=0}^{c-1}s_{j2}[B_{j2}]_c\right)\right]\int_0^\infty dr^2\left(r^2\right)^{\frac{D-2}{2}}\prod_{j1=c}^{n-2}\int_0^1 dt_{j1} \left[t_{j1}(1-t_{j1})\right]^{\frac{D-3-j1}{2}}\nonumber \\ 
&\times exp\left[-\left(s_{n-1}(r^2+m_{n-1}^2)+\sum_{j1=c}^{n-2}s_{j1}[B_{j1}]_c\left(t_{j1}-[T_{j1}]_c\right)\right)\right] ,
\end{align}
Now for $j_1\geq c$  the $[B_{j1}]_c$ are not independent of $k^E$ (the loop momentum) since in the formula of $[B_{j1}]_c$, eq.(4.4) of \cite{Abreu:2017ptx}, for $j1<c$ we have $Y_{j1}$ which demands all propagators with $j<j1$ to be cut. As this is not true for the $j\geq c$ this formula is not valid and we use the usual formula eq.(4.4) of \cite{Abreu:2017ptx} to represent the propagator. Now since we have $k^E$ in the integrand we have to go back to the original co-ordinate system consisting of $k^E$ as the integration variable. For this we first go back from $t_j$ to $\theta_j$. After doing all these changes we get
\begin{align}\label{alphacut13}
&C_cI_n=(-1)^n\frac{2^{\sum_{j=0}^{c-1}(D-2-j)}e^{\gamma_{E}\epsilon}}{\pi^{\frac{n-1}{2}}\Gamma\left(\frac{D-n+1}{2}\right)\Gamma(n)}\prod_{j=0}^{c-1} \left[t_{j,p}(1-t_{j,p})\right]^{\frac{D-3-j}{2}}\prod_{k=0}^{n-1}\int_0^\infty ds_{k}\hspace{.1cm} exp\left[-\left(\sum_{j2=0}^{c-1}s_{j2}[B_{j2}]_c\right)\right]
\nonumber\\
&\times \int_0^\infty dr^2\left(r^2\right)^{\frac{D-2}{2}}\prod_{j1=c}^{n-2}\int_0^{\pi} d\theta_{j1} \left[\sin{\theta_{j1}}\right]^{D-2-j1} exp\left[-\left(\sum_{j1=c}^{n-1}s_{j1}\left((k^E-q_{j1}^E)^2-m_{j1}^2\right)\right)\right]   ,
\end{align}
Contrary to the original set of variables where we had n vaiables here we only have n-c variables with the remaining c variables $(\theta_j,0\leq j\leq c-1)$ now being constant. If we define a vector $k'$ given by
\begin{align}\label{alphacut14}
    k'=r\prod_{j=0}^{c-1}\sin{\theta_j}\left(\cos{\theta_c},\cos{\theta_{c+1}}\sin{\theta_c},...,\cos{\theta_{n-2}}\prod_{j=c}^{n-2}\sin\theta_j,\mathbf{1}_{D-n+1}\,\prod_{j=c}^{n-2}\sin\theta_j\right)
\end{align}
the differential volume for this D-c dimensional vector is given by
\begin{align}\label{alphacut15}
    \int d^{D-c}k'=\frac{i\pi^{\frac{D-n+1}{2}}}{\Gamma\left(\frac{D-n+1}{2}\right)}\prod_{j=0}^{c-1}[\sin{\theta_j}]^{D-c}\int_{0}^{\infty}dr\, r^{D-c-1}\int_{0}^{\pi}\prod_{j=c}^{n-2}d\theta_{j}\,   \left[\sin{\theta_{j}}\right]^{D-2-j}
\end{align}
So substituting in eq.(\ref{alphacut13}) we get
\begin{align}\label{alphacut16}
&C_cI_n=(-1)^n\frac{2^{\sum_{j=0}^{c-1}(D-2-j)}e^{\gamma_{E}\epsilon}}{\pi^{\frac{D}{2}}\Gamma(n)}\prod_{j=0}^{c-1} \left[t_{j,p}(1-t_{j,p})\right]^{\frac{D-3-j}{2}}\prod_{k=0}^{n-1}\int_0^\infty ds_{k}\hspace{.1cm} exp\left[-\left(\sum_{j2=0}^{c-1}s_{j2}[B_{j2}]_c\right)\right]
\nonumber\\
&\times \prod_{j=0}^{c-1}[\sin{\theta_j}]^{c-D} \int d^{D-c}k' \,r^c  exp\left[-\left(\sum_{j1=c}^{n-1}s_{j1}\left((k^E-q_{j1}^E)^2-m_{j1}^2\right)\right)\right]   ,
\end{align}
We are now successful in writing the integration variables in a way that the new co-ordinate system is similar in representation to the original $k^E$ one but it is still not exactly the same as $k'\neq k^E$. For making the variables exactly equal we will define one more $c$ dimensional vector given by
\begin{align}\label{alphacut17}
    k''=r\left(\cos{\theta_0},\cos{\theta_{1}}\sin{\theta_0},...,\cos{\theta_{c-1}}\prod_{j=0}^{c-2}\sin\theta_j\right)
\end{align}
This vector is perpendicular to $k'$, has the angular part constant and summing it together with $k'$ gives $k^E$
\begin{align}\label{alphacut18}
    k^E=k'+k''
\end{align}
Also we have 
\begin{align}\label{alphacut19}
    (k^E)^2=(k')^2+(k'')^2=r^2=r^2X^2+r^2Y^2=r^2X^2\left(1+\frac{Y^2}{X^2}\right)=\lvert k'\rvert^2\left(1+\frac{Y^2}{X^2}\right)
\end{align}
where 
\begin{align}\label{alphacut20}
    r X=\lvert k' \rvert, \hspace{.3cm} r Y=\lvert k'' \rvert
\end{align}
with
\begin{align}\label{alphacut21}
    X^2=\prod_{j=0}^{c-1}[\sin{\theta_j}]^2,\hspace{.3cm}
    Y^2=\sum_{j=0}^{c-1}\left([\cos{\theta_j}]^2\prod_{k=0}^{j-1}[\sin{\theta_k}]^2\right), 
\end{align}
So the argument of the exponential in eq.(\ref{alphacut16}) becomes
\begin{align}\label{alphacut22}
    &\sum_{j=c}^{n-1}s_{j}\left((k^E-q_{j}^E)^2-m_{j}^2\right)=(k^E)^2\left(\sum_{j=c}^{n-1}s_{j}\right)-2k^E\cdot \left(\sum_{j=c}^{n-1}s_{j}q_j^E\right) +\sum_{j=c}^{n-1}s_{j}((q_j^E)^2 +m_j^2)\nonumber\\
    &=\lvert k'\rvert^2\left(1+\frac{Y^2}{X^2}\right)\left(\sum_{j=c}^{n-1}s_{j}\right)-2k''\cdot \left(\sum_{j=c}^{n-1}s_{j}q_j^E\right)-2k'\cdot \left(\sum_{j=c}^{n-1}s_{j}q_j^E\right)+\sum_{j=c}^{n-1}s_{j}((q_j^E)^2 +m_j^2)\nonumber\\
    &=\nonumber \\
    &\lvert k'\rvert^2\left(1+\frac{Y^2}{X^2}\right)\left(\sum_{j=c}^{n-1}s_{j}\right)-2\lvert k'\rvert\, Z \left(1+\frac{Y^2}{X^2}\right)^{1/2}-2\lvert k'\rvert \left\lvert\sum_{j=c}^{n-1}s_{j}q_j^E\right\rvert \cos{\alpha_0}+\sum_{j=c}^{n-1}s_{j}((q_j^E)^2 +m_j^2)
\end{align}
with Z being a constant. Using eq.(\ref{alphacut17}) and eq.(4.1) of \cite{Abreu:2017ptx} it is given by
\begin{align}\label{alphacut23}
    Z=\sum_{j=c}^{n-1}s_{j}\sum_{i=0}^{c-1}\left(q_{ji}^E\cos{\theta_i}\prod_{k=0}^{i-1}\sin{\theta_k}\right)
\end{align}
In the last step in eq.(\ref{alphacut22}) we have used the definition of dot product for the third term to write it as a product of modulii of the vectors participating in the dot product and the angle $\alpha_0$ between them. Now using all the substitutions we can write eq.(\ref{alphacut16}) as
\begin{align}\label{alphacut24}
&C_cI_n=(-1)^n\frac{2^{\sum_{j=0}^{c-1}(D-2-j)}e^{\gamma_{E}\epsilon}}{\pi^{\frac{D}{2}}\Gamma(n)}\prod_{j=0}^{c-1} \left[t_{j,p}(1-t_{j,p})\right]^{\frac{D-3-j}{2}}X^{c-D}\left(1+\frac{Y^2}{X^2}\right)^{c/2}
\nonumber\\
&\times  \prod_{k=0}^{n-1}\int_0^\infty ds_{k}\hspace{.1cm} exp\left[-\left(\sum_{j=c}^{n-1}s_{j}((q_j^E)^2 +m_j^2)+\sum_{j2=0}^{c-1}s_{j2}[B_{j2}]_c\right)\right]\int d^{D-c}k' \,\lvert k' \rvert^c  \nonumber \\
&\times exp\left[-\left(\lvert k'\rvert^2\left(1+\frac{Y^2}{X^2}\right)\left(\sum_{j=c}^{n-1}s_{j}\right)-2\lvert k'\rvert\, Z \left(1+\frac{Y^2}{X^2}\right)^{1/2}-2\lvert k'\rvert \left\lvert\sum_{j=c}^{n-1}s_{j}q_j^E\right\rvert \cos{\alpha_0}\right)\right],
\end{align}
Here $X$ and $Y$ are constants with respect to integration varaibles and hence we have taken them out of the integration. Now baring the new variable $\alpha_0$ we can see that the integration is entirely in terms of $k'$ variable. To do this integration we will again go back to the spherical co-ordinate system of $k'$ but now we will re-orient $k'$ such that the angle between $k'$ and $\sum_{j=c}^{n-1}s_{j}q_j^E$ is $\alpha_0$. To do this we define the new co-ordinate system such that 
\begin{align}\label{alphacut25}
    &\sum_{j=c}^{n-1}s_{j}q_j^E=\left(\left\lvert\sum_{j=c}^{n-1}s_{j}q_j^E\right\rvert, \mathbf{0}_{D-c-1}\right)\hspace{1cm} \text{and} \nonumber\\
    &k'=\lvert k'\rvert\left(\cos\alpha_0,\cos\alpha_1\sin\alpha_0,\ldots,\cos\alpha_{\lfloor D-c\rfloor-2}\,\prod_{j=0}^{\lfloor D-c\rfloor-3}\sin\alpha_j,\mathbf{1}_{D-c-(\lfloor D-c\rfloor)+1}\,\prod_{j=0}^{\lfloor D-c\rfloor-2}\sin\alpha_j\right)
\end{align}
So we see that the dot product of these two vectors will give rise to the required term in the argument of the exponential. Also the other terms in the exponent will not be affected since they are just products of modulus of the $k'$ vector with constants which are rotationally invariant. Now eq.(\ref{alphacut24}) can be rewritten as
\begin{align}\label{alphacut26}
&C_cI_n= \nonumber \\
&(-1)^n\frac{2^{\sum_{j=0}^{c-1}(D-2-j)}\pi^{\frac{c+(\lfloor D-c\rfloor)-1}{2}}e^{\gamma_{E}\epsilon}}{\Gamma(n)\Gamma\left(\frac{D-c-(\lfloor D-c\rfloor)+1}{2}\right)}\prod_{j=0}^{c-1} \left[t_{j,p}(1-t_{j,p})\right]^{\frac{D-3-j}{2}}X^{c-D}\left(1+\frac{Y^2}{X^2}\right)^{c/2}\prod_{k=0}^{n-1}\int_0^\infty ds_{k} \times 
\nonumber\\
& exp\left[-\left(\sum_{j=c}^{n-1}s_{j}((q_j^E)^2 +m_j^2)+\sum_{j2=0}^{c-1}s_{j2}[B_{j2}]_c\right)\right]\int_{0}^{\infty}d\lvert k' \rvert\, \lvert k' \rvert^{D-1}\prod_{j=0}^{\lfloor D-c\rfloor-2}\int_{0}^{\pi}d\alpha_{j}\,   \left[\sin{\alpha_{j}}\right]^{D-c-2-j}  \nonumber \\
&\times exp\left[-\left(\lvert k'\rvert^2\left(1+\frac{Y^2}{X^2}\right)\left(\sum_{j=c}^{n-1}s_{j}\right)-2\lvert k'\rvert\, Z \left(1+\frac{Y^2}{X^2}\right)^{1/2}-2\lvert k'\rvert \left\lvert\sum_{j=c}^{n-1}s_{j}q_j^E\right\rvert \cos{\alpha_0}\right)\right],
\end{align}
Finally we are in a stage to do the integration. First we will do the angular integrations. Here the $\alpha_0$ integration can be done in terms of hypergeometric ${}_0F_1$ functions or Bessel functions and for the remaining angular variables the integration is trivial. After integration we have
\begin{align}\label{alphacut27}
&C_cI_n=(-1)^n\frac{2^{\sum_{j=0}^{c-1}(D-2-j)}e^{\gamma_{E}\epsilon}}{\pi^{\frac{ D-1}{2}}\Gamma(n)}\prod_{j=0}^{c-1} \left[t_{j,p}(1-t_{j,p})\right]^{\frac{D-3-j}{2}}X^{c-D}\left(1+\frac{Y^2}{X^2}\right)^{c/2}\prod_{k=0}^{n-1}\int_0^\infty ds_{k}
\nonumber\\
&\times  \hspace{.1cm} exp\left[-\left(\sum_{j=c}^{n-1}s_{j}((q_j^E)^2 +m_j^2)+\sum_{j2=0}^{c-1}s_{j2}[B_{j2}]_c\right)\right]\int_{0}^{\infty}d\lvert k' \rvert\, \lvert k' \rvert^{D-1} \frac{2\pi^\frac{D-c}{2}}{\Gamma\left(\frac{D-c}{2}\right)}\hspace{.2cm} \times \nonumber \\
&  {}_0F_1\left(;\frac{D-c}{2};\lvert k'\rvert^2 \left\lvert\sum_{j=c}^{n-1}s_{j}q_j^E\right\rvert^2\right) exp\left[-\left(\lvert k'\rvert^2\left(1+\frac{Y^2}{X^2}\right)\left(\sum_{j=c}^{n-1}s_{j}\right)-2\lvert k'\rvert\, Z \left(1+\frac{Y^2}{X^2}\right)^{1/2}\right)\right],
\end{align}
After expanding the hypergeometric ${}_0F_1$ function using its definition we can integrate the $k'$ variable using hypergeometric ${}_1F_1$ functions and we have the following result
\begin{align}\label{alphacut28}
&C_cI_n=(-1)^n\frac{2^{\sum_{j=0}^{c-1}(D-2-j)}e^{\gamma_{E}\epsilon}}{\pi^{\frac{ D-1}{2}}\Gamma(n)}\prod_{j=0}^{c-1} \left[t_{j,p}(1-t_{j,p})\right]^{\frac{D-3-j}{2}}X^{c-D}\left(1+\frac{Y^2}{X^2}\right)^{\frac{c-D}{2}}\prod_{k=0}^{n-1}\int_0^\infty ds_{k} \hspace{.1cm} \pi^\frac{D-c}{2} \times
\nonumber\\
&   exp\left[-\left(\sum_{j=c}^{n-1}s_{j}((q_j^E)^2 +m_j^2)+\sum_{j2=0}^{c-1}s_{j2}[B_{j2}]_c\right)\right]\sum_{m=0}^{\infty} \frac{2\left\lvert\sum_{j=c}^{n-1}s_{j}q_j^E\right\rvert^{2m}\left(\left(1+\frac{Y^2}{X^2}\right) \left(\sum_{j=c}^{n-1}s_{j}\right)\right)^{-m}} {\left(\sum_{j=c}^{n-1}s_{j}\right)^{D/2}\Gamma(m+1)\Gamma\left(\frac{D-c+2m}{2}\right)} \nonumber \\
& \times \left[\frac{Z\hspace{.1cm} \Gamma\left(\frac{D+1+2m}{2}\right)}{\left(\sum_{j=c}^{n-1}s_{j}\right)^{1/2}}\hspace{.1cm} {}_1F_1\left(\frac{D+1+2m}{2};\frac{3}{2};\frac{Z}{\left(\sum_{j=c}^{n-1}s_{j}\right)}\right) \right.\nonumber \\ & \hspace{6cm}+ \left. \frac{ \Gamma\left(\frac{D+2m}{2}\right)}{2}\hspace{.1cm} {}_1F_1\left(\frac{D+2m}{2};\frac{1}{2};\frac{Z}{\left(\sum_{j=c}^{n-1}s_{j}\right)}\right) \right]
\end{align}
 Now if we rewrite it in a standard Schwinger parametrization form then we get the following form of the generalized U and F polynomials in the cut case.
 \begin{equation}\label{alphacut29}
     \begin{split}
     &\frac{F'(s)}{U'(s)}=  \\
     &\sum_{j=c}^{n-1}s_{j}((q_j^E)^2 +m_j^2)+\sum_{j=0}^{c-1}s_{j}[B_{j}]_c - \ln \left[ \sum_{m=0}^{\infty} \frac{2\left\lvert\sum_{j=c}^{n-1}s_{j}q_j^E\right\rvert^{2m}\left(\left(1+\frac{Y^2}{X^2}\right) \left(\sum_{j=c}^{n-1}s_{j}\right)\right)^{-m}} {\Gamma(m+1)\Gamma\left(\frac{D-c+2m}{2}\right)} \times  \right.\\
     & \left\lbrace \frac{Z\hspace{.1cm} \Gamma\left(\frac{D+1+2m}{2}\right)}{\left(\sum_{j=c}^{n-1}s_{j}\right)^{1/2}}\hspace{.1cm} {}_1F_1\left(\frac{D+1+2m}{2};\frac{3}{2};\frac{Z}{\left(\sum_{j=c}^{n-1}s_{j}\right)}\right) \right.  \\  & \hspace{6cm} \left. \left. + \frac{ \Gamma\left(\frac{D+2m}{2}\right)}{2}\hspace{.1cm} {}_1F_1\left(\frac{D+2m}{2};\frac{1}{2};\frac{Z}{\left(\sum_{j=c}^{n-1}s_{j}\right)}\right)\right\rbrace \right]
     \end{split}
 \end{equation}
 \begin{align}\label{alphacut30}
     U'(s)=\left(\sum_{j=c}^{n-1}s_{j}\right)
 \end{align}
 Here we can take $D=d-2\epsilon$ with $d$ being an integer in order to have a proper representation of divergences occuring due to $c\geq D$. Now let us check the validity of these equations by checking whether they match the uncut case when $c=0$. When $c=0$ we have $k''=0$ and hence $Y=Z=0$, also $k'=k^E$ and hence $X=1$. Inserting all this in eq.(\ref{alphacut29}) and eq.(\ref{alphacut30}) we have
 \begin{align}\label{alphacut31}
     \frac{F'(s)}{U'(s)}&=\sum_{j=0}^{n-1}s_{j}((q_j^E)^2 +m_j^2)- \ln \left[ \sum_{m=0}^{\infty} \frac{\left\lvert\sum_{j=0}^{n-1}s_{j}q_j^E\right\rvert^{2m}\left(\left(\sum_{j=0}^{n-1}s_{j}\right)\right)^{-m}} {\Gamma(m+1)}   \right] \nonumber
     \\
     &=\sum_{j=0}^{n-1}s_{j}((q_j^E)^2 +m_j^2)-   \frac{\left(\sum_{j=0}^{n-1}s_{j}q_j^E\right)^{2}} {\left(\sum_{j=0}^{n-1}s_{j}\right)} 
 \end{align}
 \begin{align}\label{alphacut32}
     U'(s)=\left(\sum_{j=0}^{n-1}s_{j}\right)
 \end{align}
 which is exactly equal to the U and F polynomial equations for the uncut integral in the one loop case. Also the cut integral becomes
\begin{align}\label{alphacut33}
&C_0I_n=\frac{(-1)^n e^{\gamma_{E}\epsilon}}{\pi^{\frac{ -1}{2}}\Gamma(n)}\prod_{k=0}^{n-1}\int_0^\infty \frac{ds_{k}}{\left(\sum_{j=0}^{n-1}s_{j}\right)^{D/2}} \hspace{.1cm} exp\left[-\left(\sum_{j=c}^{n-1}s_{j}((q_j^E)^2 +m_j^2)-   \frac{\left(\sum_{j=0}^{n-1}s_{j}q_j^E\right)^{2}} {\left(\sum_{j=0}^{n-1}s_{j}\right)}\right)\right] 
\end{align} 
 And now after some simplification we can construct the Feynman parametrisation equation for the cut integral given by
 \begin{align}\label{alphacut34}
&C_cI_n=\nonumber\\
&(-1)^n\frac{2^{\sum_{j=0}^{c-1}(D-2-j)}e^{\gamma_{E}\epsilon}}{\pi^{\frac{ D-1}{2}}\Gamma(n)}\prod_{j=0}^{c-1} \left[t_{j,p}(1-t_{j,p})\right]^{\frac{D-3-j}{2}} \Gamma\left(n-\frac{D}{2}\right)\prod_{k=0}^{n-1}\int_0^\infty d\alpha_{k} \hspace{.1cm} \frac{\delta\left(1-\sum_{j=0}^{n-1}\alpha_j\right)}{F'(\alpha)^{n-\frac{D}{2}}U'(\alpha)^{D-n}}
\end{align}
where $U'(\alpha)$ and $F'(\alpha)$ are given by eq.(\ref{alphacut29}) and eq.(\ref{alphacut30}) with $\alpha$ in place of s.

\end{document}